\documentclass[a4paper,12pt]{article}

\usepackage{jheppub} 

\usepackage{float}
\usepackage{ulem}
\usepackage{cancel}
\usepackage{enumitem}

\usepackage[latin1]{inputenc}

\usepackage{comment}

\usepackage{color}
\usepackage{graphicx}

\usepackage{shuffle}

\renewcommand{\thefootnote}{\fnsymbol{footnote}}
\renewcommand{\thanks}[1]{\footnote{#1}}

\newcommand{\bea}{\begin{eqnarray}}
\newcommand{\eea}{\end{eqnarray}}
\newcommand{\ee}{\end{equation}}
\newcommand{\be}{\begin{equation}}

\newcommand{\no}{\nonumber}

\def\eqn#1{eq.~(\ref{#1})}
\def\fig#1{fig.~{\ref{#1}}}
\def\sec#1{sec.~{\ref{#1}}}

\def\cN{{\cal N}}

\def\ha{\hat a}
\def\hb{\hat b}
\def\hc{\hat c}
\def\hd{\hat d}
\def\he{\hat e}

\title{\boldmath
Explicit Formulae for Yang-Mills-Einstein Amplitudes from the Double Copy}

\author[a]{Marco Chiodaroli,}
\author[b]{Murat G\"{u}naydin,}
\author[a,c]{Henrik Johansson,}
\author[b]{and Radu Roiban}

\affiliation[a]{Department of Physics and Astronomy, \\ Uppsala University, 75108 Uppsala, Sweden}
\affiliation[b]{Institute for Gravitation and the Cosmos, \\ The Pennsylvania State University,
University Park, PA 16802, USA}
\affiliation[c]{Nordita, Stockholm University and KTH Royal Institute of Technology,\\  Roslagstullsbacken 23,
10691 Stockholm, Sweden}

\emailAdd{marco.chiodaroli@physics.uu.se}
\emailAdd{mgunaydin@psu.edu}
\emailAdd{henrik.johansson@physics.uu.se}
\emailAdd{radu@phys.psu.edu}

\abstract{
Using the double-copy construction of Yang-Mills-Einstein  theories formulated in our earlier work, we obtain
compact presentations for single-trace Yang-Mills-Einstein tree amplitudes with up to five external gravitons and an arbitrary number of gluons. These are written as linear combinations of color-ordered Yang-Mills trees, where the coefficients are given by  color/kinematics-satisfying numerators in a Yang-Mills$\,+\,\phi^3$ theory. 
The construction outlined in this paper holds in general dimension and extends straightforwardly to supergravity theories.
For one, two, and three external gravitons, our expressions give identical or simpler presentations of amplitudes already constructed through string-theory considerations or the scattering equations formalism. 
Our results are based on color/kinematics duality and gauge invariance, and strongly hint at a recursive structure underlying the single-trace amplitudes with 
an arbitrary number of gravitons.   We also present explicit expressions for all-loop single-graviton Einstein-Yang-Mills amplitudes in terms of Yang-Mills amplitudes and, through gauge invariance, derive new all-loop amplitude relations for Yang-Mills theory. }

\preprint{UUITP-07/17 \\
\phantom{~} \hfill NORDITA-2017-19}

\begin{document}
\maketitle
\flushbottom

\renewcommand{\thefootnote}{\arabic{footnote}}

\section{Introduction}

In recent years, the double-copy construction formulated in the work of Bern, Carrasco, and one of the current 
authors (BCJ) \cite{Bern:2008qj,Bern:2010ue} has gained a central role in our understanding of perturbative quantum gravity.
Despite their apparent differences, amplitudes in gravity have been shown to be closely related to the ones of Yang-Mills (YM) theory.
At tree level, this connection was first established by the Kawai-Lewellen-Tye (KLT) relations in string theory \cite{Kawai:1985xq}.
The BCJ double copy  gives a more systematic understanding of this structure, including its extension
to loop level and to larger classes of theories, some of which might not have a string-theory origin.
It relies on a Lie-algebraic structure of 
certain kinematic building blocks in 
diagrammatic presentations of gauge-theory amplitudes.
Once gauge-theory integrands are available in a form in which their algebraic properties are manifest,
they can be rearranged to express the integrands of gravity amplitudes as double copies of the ones of gauge theory.
In particular, invariance under linearized diffeomorphisms
of gravity amplitudes is 
an immediate consequence of 
the gauge invariance of the gauge-theory amplitudes, as we discuss in sec.~\ref{revsec}. 

The initial formulation of the double-copy construction \cite{Bern:2008qj,Bern:2010ue}  has been significantly extended
to include applications to
pure supergravities with reduced supersymmetry \cite{Johansson:2014zca}, broad classes
of theories with various matter contents and interactions \cite{Carrasco:2012ca,Chiodaroli:2013upa,Chiodaroli:2015wal,Anastasiou:2016csv},
and, most recently, effective (non-gravitational) theories such as the Dirac-Born-Infeld/special galileon theory \cite{Chen:2013fya, Chen:2014dfa, Cachazo:2014xea,  Du:2016tbc,  Carrasco:2016ldy, Chen:2016zwe, Cheung:2016prv},
prompting the question of whether all gravity theories  can be ``deconstructed"
as double copies of suitably-chosen pairs of gauge theories.
The double-copy structure appears naturally in the string-theory framework \cite{Stieberger:2009hq,BjerrumBohr:2009rd,BjerrumBohr:2010zs,Mafra:2011kj}. It is also an intrinsic 
feature of modern approaches to amplitudes in quantum field theories involving gravity, 
such as the scattering equations formalism \cite{Cachazo:2013hca,Cachazo:2013gna,Cachazo:2013iea} and ambitwistor string theories
\cite{Mason:2013sva,Geyer:2014fka}. 
The double copy has been formulated at the level of off-shell linearized supermultiplets in refs. \cite{Borsten:2013bp,Anastasiou:2013hba,Anastasiou:2014qba,Anastasiou:2015vba}.
Finally, a double-copy structure relating classical solutions of gauge and gravity  theories was identified in refs.  
\cite{Monteiro:2014cda,Luna:2015paa,Luna:2016due,Cardoso:2016ngt,Cardoso:2016amd},
raising hope for its application as a solution-generating technique for asymptotically-flat perturbative solutions in general relativity \cite{Luna:2016hge}.

While a considerable amount of effort has been devoted to investigating gravities 
coupled to abelian matter  from a double-copy perspective, amplitudes in gravitational theories with non-abelian gauge interactions have been less widely studied despite 
their phenomenological relevance.
At the same time, the literature has long explored and classified the constraints imposed by introducing non-abelian
gauge interactions in supergravity, uncovering a rich variety of physical features. In theories with a  high number of supersymmetries,
non-abelian gauge interactions involving R-symmetry introduce, in general, a non-zero cosmological constant. For the 
maximally-supersymmetric theory \cite{Cremmer:1979up},
$SO(8)$ gauging in four dimensions was first studied in  \cite{deWit:1982bul}.
In five dimensions,  $SO(p,6-p)$ $(p=0,1,2,3)$  gaugings of maximal supergravity were studied in refs. \cite{Gunaydin:1984qu}  and \cite{Gunaydin:1985cu}.
Classifying all possible compact and non-compact gaugings of extended supergravities in various dimensions
is an active research area in the  supergravity literature.
For $\cN \leq 4$ supergravities with matter, it is possible to gauge a subgroup
of their global symmetry groups  while leaving the R-symmetry ungauged. Theories of this
sort are referred to as Yang-Mills-Einstein (YME) supergravities, in contrast with the ``gauged" Yang-Mills-Einstein supergravities \cite{Gunaydin:1984ak}
for which R-symmetry is also gauged.
Examples with  $\cN=2$ supersymmetry in five dimensions were first obtained
in the work of Sierra, Townsend and one of the current authors \cite{Gunaydin:1983bi,Gunaydin:1984ak,Gunaydin:1986fg},
and have been subsequently studied in various dimensions by a large body of literature.

Theories of the YME class are particularly amenable to momentum-space perturbative calculations, as they always possess Minkowski 
vacua.\footnote{This is to be contrasted with  theories in which  R-symmetry is also gauged which may not have Minkowski vacua, either 
supersymmetric or not.} 
Certain tree-level amplitudes in Einstein gravity coupled to YM theory were first studied by Bern, De Freitas and Wong in the context of the KLT 
relations \cite{Bern:1999bx}. The current authors formulated a double-copy construction for YME amplitudes in ref. \cite{Chiodaroli:2014xia}
in which one copy is a non-supersymmetric Yang-Mills-scalar theory with particular trilinear relevant couplings (YM$\,+\,\phi^3$) and the other is pure YM theory or its supersymmetric extensions.  Schematically, this is written as ${\rm YME}=({\rm YM}+\phi^3) \otimes {\rm YM}$. In terms of 
YME tree-level scattering amplitudes, the double-copy implies that
\be
M_{n, {\rm tree}}^{\rm YME}= \sum_{i \in {\rm cubic}} \frac{n_i^{{\rm YM}+\phi^3} n_i^{\rm YM}}{D_i}= A_{n, {\rm tree}}^{{\rm YM}+\phi^3} \cdot S_{\rm KLT} \cdot A_{n, {\rm tree}}^{{\rm YM}} \equiv \alpha \cdot  A_{n, {\rm tree}}^{{\rm YM}} \, ,
\ee
where summation is over all cubic tree graphs, $S_{\rm KLT}$ is the so-called KLT kernel and $\alpha= A_{n, {\rm tree}}^{{\rm YM}+\phi^3} \cdot S_{\rm KLT}$ is a column vector 
with $(n-3)!$ non-local entries (i.e. rational functions). This implies that all YME tree amplitudes are linear combinations of YM tree amplitudes. In later sections, we shall elaborate on this point and exploit it to find explicit all-multiplicity tree amplitudes.
By picking a $(n-2)!$ basis of YM amplitudes, the coefficient vector $\alpha$ can be chosen to be local, as we shall see in sec.~\ref{explicitsec}.

Amplitudes for supergravities with different amounts of supersymmetry
can be obtained with an appropriate choice for the second gauge-theory factor entering the construction. In particular,
selecting a pure $\cN=2$ super-Yang-Mills (sYM) theory, the result of the double
copy can be identified as an infinite family of YME supergravities in four and five dimensions known as the generic Jordan family.
It is possible to 
include spontaneous symmetry breaking in the double-copy framework, as explained in ref.
\cite{Chiodaroli:2015rdg}.
Shorty after their double-copy construction became available,
YME amplitudes were also obtained in the context of scattering equations \cite{Cachazo:2014nsa,Cachazo:2014xea,Du:2016wkt} and ambitwistor strings
\cite{Casali:2015vta,Adamo:2015gia}.
Additionally, string-theory techniques have been employed  to relate amplitudes involving gravitons
to the collinear limit of gluon amplitudes \cite{Stieberger:2014cea,Stieberger:2015qja,Stieberger:2015kia}.
Indeed, the fact that YME amplitudes 
can be used to test and compare different computational techniques gives additional motivation  for their study.

Despite the existence of various techniques, explicit formulae for amplitudes in YME theories are still surprisingly rare. 
The double-copy construction relies on the availability of numerators that manifestly obey
the duality between color and kinematics for at least one of the gauge theories entering the construction.
While string theory has played a fundamental role in generating duality-satisfying numerators for particular theories
\cite{Mafra:2011kj,Mafra:2014gja,Mafra:2015mja,He:2015wgf,Mafra:2015vca},
to date there is no established technique for obtaining such numerators in generic theories whose color-ordered partial amplitudes obey the BCJ amplitude 
relations. Moreover, at higher loops, their very existence is conjectural.
Scattering equations, while exhibiting a double-copy structure, are generically difficult to solve for high numbers of 
external legs. Hence, it is difficult to translate the existing constructions into explicit formulae holding for any number of external states.

Recent progress in this direction came through the work of Stieberger and Taylor,
who used open/closed string-theory  relations  to find a simple expression for the YME tree amplitudes with
one external graviton and an arbitrary number of gluons \cite{Stieberger:2016lng}.
Their results were extended by Nandan, Plefka, Schlotterer and Wen, who used the scattering equations formalism to
give explicit formulae with up to three external gravitons in the single-trace case and up to one graviton in the double-trace case \cite{Nandan:2016pya}. In a related work~\cite{delaCruz:2016gnm} it was proven that the single-trace sector of YME tree amplitudes can always be written as a linear combination of YM color-ordered tree amplitudes; a similar conclusion was reached by studying the field-theory limit of heterotic string amplitudes~\cite{Schlotterer:2016cxa}. These results agree with the 
construction of ref.~\cite{Chiodaroli:2014xia}, where it was argued that all amplitudes in YME (including multi-trace terms
 and loop amplitudes) are obtained from the $({\rm YM}+\phi^3) \otimes {\rm YM}$ double copy.

In this paper, we take a significant step towards finding explicit all-multiplicity formulae relating YME and YM tree amplitudes. 
Focusing on the single-trace sector,
we use the double-copy prescription formulated in our earlier work and some basic properties of the amplitudes
of the gauge-theory factors to obtain explicit formulae  for YME amplitudes
involving up to five external gravitons and any number of external gluons. With one external graviton,
our results reproduce the formulae by Stieberger and Taylor~\cite{Stieberger:2016lng}. With two or three external gravitons, we find very compact
expressions which are non-trivially equivalent to the ones obtained with scattering equations techniques  \cite{Nandan:2016pya}.
The new results are a direct application of the construction in ref. \cite{Chiodaroli:2014xia} and demonstrate the power of the double-copy approach to obtaining YME amplitudes.
At the same time, we uncover additional structure which is not a priori expected:
the explicit formulae we find strongly hint at a recursive structure, and raise the
hope that a closed-form expression for any number of external gravitons might be within reach.

The remainder of the paper is organized as follows. In \sec{revsec}, we review the salient features
of the double-copy construction for YME amplitudes.
In \sec{explicitsec}, we use some basic properties of gauge-theory amplitudes,
such as the existence of a Del Duca-Dixon-Maltoni multiperipheral representation,
to pose strong constraints on YME amplitudes and show that most of their terms are dictated by gauge invariance.
We give explicit formulae for amplitudes with up to five external gravitons.
Finally, in sec. \ref{loopsec} and \ref{other}, we discuss extensions to loop level and other theories and conclude in  \sec{discussion} 
discussing several open problems. 
In the appendix, we include novel presentations of 
the BCJ amplitudes relations, which are useful in the main part of the text.

\vfill
\eject 

\section{Double-copy construction for YME amplitudes\label{revsec}}

In this section, we first argue that the double copy of scattering amplitudes of two gauge theories that obey color/kinematics duality always leads to the scattering amplitudes of 
some theory of gravity (i.e. invariant under diffeomorphisms).  
Note that the ideas presented here have been partially addressed in various contexts within 
the double-copy literature \cite{Bern:2008qj,Bern:2010ue,Bern:2010yg,Johansson:2014zca,Chiodaroli:2014xia}; here we give a more complete argument (see also
refs. \cite{Chiodaroli:2016jqw,Arkani-Hamed:2016rak}).
We conclude the section by  reviewing the construction of the amplitudes of certain classes of  
YME (super)gravity theories.    

\subsection{Two color/kinematics-dual gauge theories always gravitate \label{gaugeinv}}

$L$-loop $n$-point scattering amplitudes in any matter-coupled gauge theory can be organized as
\begin{equation}
   {\cal A}^{(L)}_{n} = i^{L-1} g^{n-2+2L} \sum_{i \in \text{cubic}}\, \int \frac{d^{LD}\ell}{(2\pi)^{LD}} \frac{1}{S_i} \frac{c_{i} n_i}{D_i} \, ,
   \label{gaugepresentation}
 \end{equation}
 where the sum runs over the $L$-loop $n$-point cubic graphs. $D_i$ stands for the product of 
the  inverse scalar propagators associated to the $i$-th graph, $S_i$ are symmetry factors, while $c_i$ 
and $n_i$ are group-theory and kinematic factors associated  with that graph.
 The latter are polynomials in scalar products of momenta, polarization
 vectors of external gluons, external spinors and flavor structure of any matter particles.  The defining commutation relations of the gauge group as well as its Jacobi identities imply that there exist  triplets of graphs
 $\{i,j,k\}$ such that $c_i - c_j=c_k$. 
A scattering amplitude is said to obey color/kinematics duality if, whenever the color-factor relations are required by gauge invariance, the kinematic numerator factors obey the same algebraic relations:
\begin{equation}
      n_i - n_j = n_k  \quad \Leftrightarrow \quad  c_i - c_j=c_k \, .
\label{duality}
\end{equation}
The color Jacobi relations imply that the kinematic numerators are not unique but can be shifted such that, without changing the amplitude, a set of numerators not obeying the relations \eqref{duality} is mapped to a set that does. For theories where all particles transform in the adjoint, the existence of such a transformation is guaranteed at tree level if the color-ordered partial amplitudes satisfy the BCJ relations~\cite{Bern:2008qj}.

Given two gauge-theory amplitudes organized as in \eqn{gaugepresentation}, with at least one of them obeying color/kinematics duality manifestly, 
the  gravity amplitudes from the double copy are obtained  
by replacing the color factors of one amplitude with the numerator factors of the other \cite{Bern:2008qj,Bern:2010ue},\footnote{Relative to previous literature, we choose to normalize the $n_i$ in a form that is more convenient when using explicit polarization vectors. We absorb a factor of $\sqrt{2}$ in $n_i$ for each cubic vertex, which changes the usual $\kappa/2$ factor in \eqn{DCformula} to a $\kappa/4$ factor. With this normalization, the color factors in \eqn{gaugepresentation} are products of $(i f^{\hat a \hat b \hat c})$'s, one for each vertex.}
\begin{equation}
{\cal M}^{(L)}_{n} = i^{L-1}\;\!\Big(\frac{\kappa}{4}\Big)^{n-2+2L} \sum_{i \in \text{cubic}}\, \int \frac{d^{LD}\ell}{(2\pi)^{LD}} \frac{1}{S_i} \frac{n_i \tilde{n}_i}{D_i} \,.
\label{DCformula}
\end{equation}
We shall show that the amplitudes given by this procedure are indeed those of some gravity theory, i.e. they are invariant under linearized diffeomorphisms. 
Invariance under linearized diffeomorphisms implies that they can follow from a fully diffeomorphism-invariant action, since only the linear part of nonlinear transformations acts on scattering amplitudes with generic (non-soft) momenta \cite{Roiban:2017iqg} (see also ref.~\cite{Sleight:2016xqq}).

We start by discussing the properties of the numerators $n_i$ 
that follow from the gauge invariance of the gauge-theory amplitudes. 
Under a linearized gauge transformation acting on a single external gluon with momentum $p$, its polarization vector becomes 
 $\varepsilon_\mu(p) \rightarrow \varepsilon_\mu(p)+ p_\mu$. 
Gauge invariance of the amplitude implies that
\begin{eqnarray}
&& n_i \rightarrow  n_i + \delta_i \, , \qquad  \delta_i = n_i \Big|_{\varepsilon\rightarrow p} \, ,
\\ 
&&  \sum_{i \in \text{cubic}}\,  \frac{c_{i} \delta_i}{D_i} = 0 \, .\label{gaugevariation}
\end{eqnarray}
In an explicit calculation, the vanishing of the second line above relies only on the explicit expressions of $\delta_i$ 
and on the algebraic properties of the color factors $c_i$.\footnote{We note that, if color/kinematics duality is manifest for any choice of transverse polarization vectors, we have sets of tree-term identities 
$\delta_i- \delta_j =\delta_k$
as a particular case of (\ref{duality}).
}

In a generic gravitational (i.e. diffeomorphism-invariant) theory, the gauge symmetry can be used to choose the off-shell graviton field to be transverse.
For the on-shell asymptotic states, the same symmetry can be used to impose simultaneously transversality and tracelessness. Consequently, the polarization tensor obeys $\varepsilon_{\mu \nu}(p) p^{\nu}_{\vphantom{\mu\nu}} = 0 = \varepsilon_{\mu \nu}(p)\eta^{\mu\nu}$. 
Amplitudes are invariant under the subset of linearized diffeomorphisms that do not modify this gauge choice. Such transformations act on the graviton polarization tensors as
\begin{equation}
\varepsilon{}_{\mu\nu}(p)  \rightarrow \varepsilon{}_{\mu\nu} (p) + p_{(\mu} q_{\nu)} \,,
\label{gaugediffeo}
\end{equation}
with both transversality and tracelessness requiring that  the arbitrary vector $q$ obeys $p \cdot q =0$.

In the double-copy framework, the graviton polarization tensor is given by the  symmetric-traceless tensor product of the polarization vectors of 
two gauge-theory gluons,  $\varepsilon_{\mu \nu}= \varepsilon_{((\mu} \tilde \varepsilon_{\nu ))}$, where the double brackets denote the symmetric-traceless part. 
Similarly, the antisymmetric part and trace part are identified with $B_{\mu\nu}$ and the dilaton, respectively. It is easy to see that the transformation \eqref{gaugediffeo} is given by the linearized gauge transformation of this product. 
Transversality and 
tracelessness of the transformed polarization tensor are  consequences of the transversality of the two gluon polarization vectors. 
We may realize the transformation \eqref{gaugediffeo} by replacing the transformed gluon polarization vector by $q$.

We now consider tree-level double-copy amplitudes. For the sake of generality, we take a set of duality-satisfying numerators 
$n_i$ only for one of the two gauge theories, and write the other set of numerators as
\begin{equation}
\tilde n^\text{\cancel{BCJ}}_i = \tilde n_i + \tilde \Delta_i \, , \qquad
\sum_{i \in \text{cubic}}  {\tilde \Delta_i c_i \over D_i} = 0 \, , \label{relGen}
\end{equation}
where $n^\text{\cancel{BCJ}}_i$ violate and $\tilde n_i$ satisfy the duality, and $\tilde \Delta_i$ are usually referred to as generalized gauge transformations. Hence, we are assuming that a presentation of the amplitude in which the duality is satisfied exists also for the second gauge theory (even though it might not be directly available).  
The second equality above stems from the fact that the transformation leaves the gauge-theory amplitude invariant, and, once more, holds due to the algebraic relations satisfied by the color factors $c_i$.

We should emphasize that here the shifts 
$\tilde \Delta_i$  
are not the result of linearized gauge transformations.
Rather, they may be interpreted as the result of gauge transformations and field redefinitions of time-ordered Green's functions before the LSZ reduction,  along the lines of ref. \cite{Lee:2015upy}.
The amplitude from the formula (\ref{DCformula}) can then be expressed as
\begin{equation}
{\cal M} = \sum_{i \in \text{cubic}} {n_i \tilde n_i^\text{\cancel{BCJ}}  \over D_i} = \sum_{i \in \text{cubic}} {n_i \tilde n_i  \over D_i} + \sum_{i \in \text{cubic}} {n_i \tilde \Delta_i  \over D_i} = \sum_{i \in \text{cubic}} {n_i \tilde n_i  \over D_i} \, ,
\label{DCformula2}
\end{equation}
where the last equality follows from (\ref{relGen}) and the fact that the numerator factors $n_i$ enjoy the same algebraic properties as the color factors $c_i$. In the above equation, we have omitted overall powers of the coupling $\kappa$.
Using (\ref{DCformula2}), we can express the variation of the double-copy amplitude at tree level under (\ref{gaugediffeo}) as
\begin{equation}
{\cal M } \rightarrow {\cal M} +  \sum_{i \in \text{cubic}} { \delta_i \ \tilde n_i \big|_{\tilde \varepsilon\rightarrow q} \over D_i}
+ \sum_{i \in \text{cubic}} { n_i \big|_{\varepsilon\rightarrow q} \tilde \delta_i \over D_i} \, .
\label{Mvariation}
\end{equation} 
The last two terms vanish because of the relation (\ref{gaugevariation}) together with the fact that the numerator factors $n_i$ and $\tilde n_i$ have 
the same algebraic properties as the color factors. 

Diffeomorphism invariance of the amplitudes at loop level follows from the invariance of the trees through generalized unitarity.\footnote{
If present, diffeomorphism anomalies may lead, in the context of generalized unitarity, to unphysical factorization properties of loop 
amplitudes~\cite{Huang:2013vha} rather than to the more familiar unitarity violation that appears in a Feynman rule calculation.} Furthermore, since the double copy makes manifest the pole and numerator structure, the factorization properties and  unitarity of the amplitude are inherited from those of the two gauge theories. Because of the sum over all cubic diagrams, crossing symmetry is also inherited from the underlying gauge theories. 
Hence, the amplitudes from the double-copy formula are, by construction, invariant under linearized diffeomorphisms and satisfy the standard field-theory properties. Having established that the double copy gives the amplitudes of 
\textit{some} gravitational theory, we now review the detailed construction for particular YME theories.

\subsection{Double-copy Maxwell- and YME (super)gravities}

In general, the spectrum does not uniquely specify the interactions of a field theory. Given its scattering amplitudes, obtained through the 
double-copy or any other construction, the Lagrangian of the theory under consideration can only be constructed by analyzing amplitudes of all multiplicities and 
extracting all higher-order interaction terms.
For sufficiently symmetric cases, however, a few interaction terms and the spectrum are sufficient to completely specify the theory. This is the 
case for the ${\cal N}=4$ Maxwell-Einstein and YME supergravity theories and the ${\cal N}=2$ Maxwell-Einstein and 
YME supergravity theories which descend from five dimensions. 

From a double-copy perspective, one gauge-theory factor is the non-supersymmetric 
YM$\,+\,\phi^3$ theory with Lagrangian  \cite{Chiodaroli:2014xia}
\bea
{\cal L}_{{\rm YM}+\phi^3} &=&-\frac{1}{4}F_{\mu\nu}^{\ha}F^{\mu\nu \ha}+\frac{1}{2}(D_\mu\phi^{A})^{\ha} (D^\mu\phi^{A})^{\ha}
 - \frac{g^2}{4} f^{\ha \hb \he} f^{\he \hc \hd}\phi^{A \ha}\phi^{B \hb}\phi^{A \hc}\phi^{B \hd}
\no \\
&&
\null + \frac{1}{3!}\lambda g F^{ABC} f^{\ha \hb \hc} \phi^{A \ha}\phi^{B \hb}\phi^{C \hc} \,.
\label{YMscalar}
\eea
Hatted indices $\ha, \hb$ run over the adjoint representation of the gauge group. Scalar fields carry additional
indices $A,B,C = 1,2,\ldots, n$. Field strength and covariant derivative are defined as
\bea F_{\mu\nu}^{\hat a}&=& \partial_\mu A_\nu^{\hat a} - \partial_\nu A_\mu^{\hat a} 
+ g  f^{\hat a \hat b \hat c}A^{\hat b}_\mu A^{\hat c}_\nu \no \ ,\\
(D_\mu\phi^{A})^{\hat a} &=& \partial_\mu \phi^{ { A \hat a}}  
+ g  f^{\hat a \hat b \hat c}A^{\hat b}_\mu \phi^{{ A} \hat c}\,. \eea

 As shown in ref. \cite{Chiodaroli:2014xia}, the requirement that four-scalar amplitudes 
from this Lagrangian obey the duality between  color and kinematics forces the constant $F^{ABC}$-tensors to obey  Jacobi relations.
Together with the reality of the scalar fields, this implies that the Lagrangian \eqref{YMscalar} is invariant under some flavor group $G$ whose adjoint representation has dimension less or equal to $n$.\footnote{Note that the $\phi^4$ term in the Lagrangian (\ref{YMscalar}) only contributes to multi-trace amplitudes at tree level~\cite{Chiodaroli:2014xia}, where the trace is with respect to the group $G$.}
It can be checked explicitly that color/kinematics duality does not impose additional constraints. 

We note that the Lagrangian (\ref{YMscalar}) does not have a straightforward supersymmetric extension. Thus, if the desired result 
is a supergravity theory and we want supersymmetry to be manifest in the construction,  the second gauge theory entering the double copy must carry the entire supersymmetry information.
There are several options for this second gauge theory, leading to different gravitational theories:
pure sYM theories with $\cN=1,2,4$ supersymmetry in four dimensions (or their higher-dimensional counterparts), as well as pure YM theory in $D$ dimensions or
its dimensional reductions (denoted as $\text{YM}_\text{DR}$).
In this paper, we focus only on constructions involving purely-adjoint theories  which we summarize in table \ref{TheoryConstructions1}.
Extensions of the double-copy construction to gauge theories which possess fields in matter (non-adjoint) representations
have been studied in refs. \cite{Chiodaroli:2013upa,Johansson:2014zca,Chiodaroli:2015rdg,Chiodaroli:2015wal}
(see also \cite{Chiodaroli:2016jqw} for a short review).

\begin{table}[t]
	\centering
	\begin{tabular}{|l||c|c|}
		\hline
		Gravity coupled to YM & Gauge theory 1  &  Gauge theory 2
		\\
		\hline \\[-13pt]
		\hline
		${\cal N}=4$ YMESG theory  & YM~+~$\phi^3$ & ${\cal N}=4$ sYM
		\\
		${\cal N}=2$ YMESG theory (gen.Jordan)  & YM~+~$\phi^3$ & ${\cal N}=2$ sYM
		\\
		${\cal N}=1$ YMESG theory & YM~+~$\phi^3$  & ${\cal N}=1$ sYM
		\\
		${\cal N}=0$ YME + dilaton + $B^{\mu\nu}$  & YM~+~$\phi^3$ &  YM
		\\
		${\cal N}=0$ ${\rm YM}_{\rm DR}$-E + dilaton + $B^{\mu\nu}$  & YM~+~$\phi^3$ &  ${\rm YM}_{\rm DR}$
		\\
		\hline
	\end{tabular}
	\small \caption[a]{\small
		Amplitudes in YME gravity theories for different number of supersymmetries, corresponding to different choices for the left gauge-theory
		factor entering the double copy.
		\label{TheoryConstructions1}}
\end{table}

\begin{enumerate}
	
 \item {\bf $\cN=4$ YME supergravities}.
 $\cN=4$ supergravity can only be coupled to $\cN=4$ vector multiplets and the global symmetry group of five dimensional  $\cN=4$ 
 Maxwell-Einstein supergravity with $n$ vector multiplets is fixed by supersymmetry to be $SO(5,n)\times SO(1,1)$.  Its R-symmetry group 
 is $USp(4)\equiv Spin(5)$. Gauging a subgroup $K$ of the $SO(n)$ symmetry of these theories leads to $\cN=4$ YME theories. The gauging 
 does introduce a potential for the scalar fields in these theories, but they admit Minkowski vacua. 
The bosonic part of these YME theories was given in  ref.~\cite{Chiodaroli:2015rdg} following ref.~\cite{DallAgata:2001wgl}.
To construct the amplitudes of these $\cN=4$ YME theories, 
 one uses  the $\cN=4$ sYM theory for the second set of numerators \cite{Chiodaroli:2015rdg} . 
This is the maximum amount
of manifest supersymmetry allowed in our construction since the  YM$\,+\,\phi^3$ theory
does not have straightforward supersymmetric extensions.
A different (and more involved) construction would be required for reproducing
the amplitudes of supergravities with gauged R-symmetry. 

\item {\bf $\cN=2$ YME supergravities}. 
This is the most intensely studied case of the construction. The second gauge-theory factor entering 
the double copy is the pure $\cN=2$ sYM theory.
$\cN=2$ YME theories which admit an uplift to five dimensions are known
very explicitly \cite{Gunaydin:1984ak,Gunaydin:1985cu,Gunaydin:1986fg}.
Their bosonic sector is  given by the five-dimensional Lagrangian
\bea
 \label{gaugedL}
e^{-1} {\cal L} &=& -{R \over 2} -{1\over 4} \text{\textit{\aa}}_{IJ} \mathcal{F}^I_{\mu \nu} \mathcal{F}^{J \mu \nu }
- {1 \over 2} g_{xy} \mathcal{D}_\mu \varphi^x \mathcal{D}^\mu \varphi^y +  \no
{e^{-1} \over 6\sqrt{6}} C_{IJK} \epsilon^{\mu\nu\rho\sigma\lambda} \left\{ \vphantom{1\over 2}
 F^I_{\mu \nu} F^J_{\rho \sigma} A^K_{\lambda} \right. \\
&& \left.  + {3 \over 2} g f^{K}{}_{ J' K'} F^I_{\mu \nu} A^J_\rho A^{J'}_\sigma A^{K'}_\lambda +
{3\over 5} g^2 A^I_\mu f^{J}{}_{ I' J'} A^{I'}_\nu A^{J'}_\rho f^{K}{}_{K' L'} A^{K'}_\sigma A^{L'}_{\lambda} \right\}  .
\label{YMESGLag}\eea
The indices $I,I',J,J',\ldots=0,1,\ldots,  n$ and $x,y=1,\ldots,  n$ run over the vectors and scalars.
$\mathcal{F}^I_{\mu \nu}$ and $\mathcal{D}_\mu$ denote covariant field strengths and covariant derivatives
and $f^{IJK}$ are the gauge-group structure constants.
As explained in ref. \cite{Gunaydin:1983bi}, such theories are completely
specified by the  symmetric tensors $C_{IJK}$ in \eqn{YMESGLag} together with
the choice of gauge group.
By computing three-point amplitudes, it is possible to read off the $C_{IJK}$ tensor and identify the theories from the double copy as a well-known family of supergravities referred to in the literature as the generic Jordan family. We should note that $\cN=2$ truncations of five-dimensional $\cN=4$ Maxwell-Einstein supergravity theories belong to the generic Jordan family~\cite{Chiodaroli:2015rdg}. 

\item {\bf $\cN=1$ YME theories}. In this case, we use the numerators from pure  $\cN=1$ sYM. The resulting YME supergravity can be seen as a truncation of the $\cN=2$ case.

\item {\bf non-supersymmetric YME theories}. A non-supersymmetric choice for the second gauge theory leads to a $\cN=0$ YME theory.
In this paper, we shall
focus on the simplest case of the construction and take a pure YM theory as one of the gauge-theory factors, using a double-copy of the form
\begin{equation}
 \text{YME} = (\text{YM} + \phi^3) \otimes \text{YM}  \, .
\end{equation}
The spectrum of the theory includes the graviton, an appropriate number of gluons, a dilaton, and a two-form field. In principle, amplitude contributions from dilaton
and  two-form field can be removed by introducing ghost fields in the double copy for loop amplitudes, as outlined in ref. \cite{Johansson:2014zca}.
Since the non-supersymmetric YM theory can be  regarded as a truncation of pure $\cN=2$ sYM theory,
it is possible to obtain its Lagrangian by truncating (\ref{YMESGLag}).
\end{enumerate}
For later
reference, we give the asymptotic states of the $\cN=2,0$ YME theory in terms of tensor product of the states of the two gauge theories:
\begin{eqnarray} 
\cN=2 \text{ YME}: &&
\left\{ \begin{array}{lcr}
h_{--} = A_- \otimes A_- \  & & \no \\
A^{-1}_- =  \bar \phi \otimes A_- \   &  \;\; \;\; \; A^0_- = \phi \otimes A_-  \;\;\;\;   &  A^A_- = A_- \otimes \phi^A  \  \no \\ 
\;\, i \bar z^0 = A_+ \otimes A_-  \   &   i \bar z^A = \bar \phi \otimes \phi^A  & \  
\end{array} \right. \, , \\
\cN=0 \text{ YME}: &&
\left\{ \begin{array}{c} 
h_{--} = A_- \otimes A_-  \\
\, A^A_- = A_- \otimes \phi^A \\
\;\, i \bar z^0 = A_+ \otimes A_-
\end{array} \right.
\label{map1} \, ,
\end{eqnarray}
where the CPT-conjugate states are not shown explicitly. The supergravity gauge coupling constant $g_s$ is related to the parameter $\lambda$ in (\ref{YMscalar}) as 
\begin{equation}
g_s= \left( {\kappa \over 4} \right) \lambda  \, .
\end{equation}

Finally, while this paper will focus on the unbroken-gauge
phase of the YME theories, the investigation
of spontaneously-broken YME theories in the double-copy framework has been initiated in ref. \cite{Chiodaroli:2015rdg}.

\section{Explicit YME amplitudes\label{explicitsec}}

A first key ingredient in our construction is that, for the double-copy prescription to
lead to sensible gravity amplitudes, 
it is sufficient that only one of the two sets of gauge-theory numerators obey the duality manifestly. For the considerations in this paper, an advantageous choice is to make the numerators of the YM$\,+\,\phi^3$ theory obey the duality, since it is very simple to work with the scalar sector of theory.
Furthermore, we can exploit the fact that the numerators can be put in a $(n-2)!$ basis using the kinematic Jacobi relations, as it can be done for the color factors. In particular, choosing the Kleiss-Kuijf basis \cite{Kleiss:1988ne} leads to the Del Duca-Dixon-Maltoni (DDM) decomposition of gauge-theory amplitudes \cite{DelDuca:1999rs}.

\subsection{DDM decomposition and YM$\,+\,\phi^3$ trees}
\begin{figure}
\begin{center}
\includegraphics[width=0.65\textwidth]{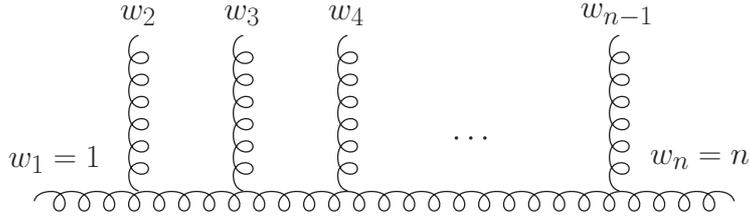}
\caption{A multiperipheral (or half-ladder) graph for YM theory. The particles are labeled by the word $w$, where the first and last element are kept fixed.  \label{figmulti}}
\end{center}
\end{figure}

An illustrative starting point is to consider the tree amplitudes in YM theory written in the DDM form \cite{DelDuca:1999rs},
\bea
{\cal A}^\text{YM}_n (1 , \ldots ,  n ) &=& -i  g^{n-2} \sum_{i \in \text{cubic}} {c_i  \, n_i^{\rm YM} \over D_i}  \\ &=& -i g^{n-2} \!\!
\sum_{ w \in S_{n-2}} C^{\rm DDM}(1,w_2, \ldots, w_{n-1},n) \, A^\text{YM}_n
(1, w_2 , \ldots , w_{n-1}, n)\,, \no  \label{DDM}
\eea
where $A^\text{YM}_n(w)$ are the color-ordered partial tree amplitudes and
the sum runs over all $(n-2)!$ words $w$ that label the different ``multiperipheral" (or ``half-ladder'') graphs with $w_1=1$ and $w_n=n$ fixed (see \fig{figmulti}). The corresponding color factors are 
\be
C^\text{DDM}(w) = i^{n-2} f^{\hat a_{w_1} \hat a_{w_2} \hat x_1} f^{\hat x_1  \hat a_{w_3} \hat x_2} \cdots f^{\hat x_{n-3} \hat  a_{w_{n-1}}\hat a_{w_{n}}} \, .
\ee

Our goal is then to replace the color factors appearing in the DDM form with duality-satisfying numerators of the YM$\,+\,\phi^3$ theory given in \eqn{YMscalar}. Thanks to the DDM choice, we need only to specify the numerators that belong to multiperipheral graphs. However, since the numerators obey color/kinematics duality, the remaining non-multiperipheral numerators can be in principle obtained through kinematic Jacobi relations from the multiperipheral ones.  

\begin{figure}
	\begin{center}
		\includegraphics[width=0.96\textwidth]{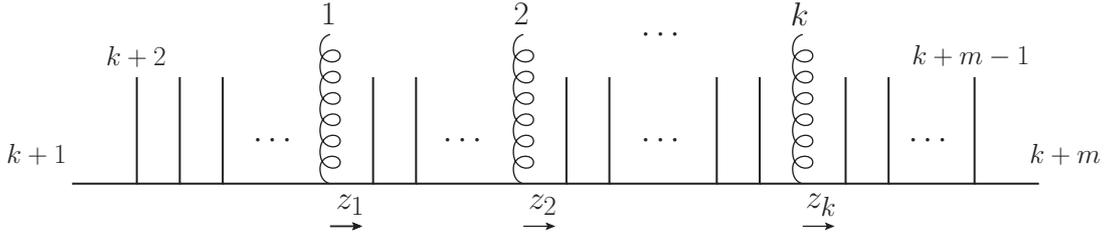}
		\caption{A typical multiperipheral (or half-ladder) graph for the YM+$\phi^3$ theory. The gluons are labeled as $1,2,\ldots, k$ and the remaining $m$ particles are canonically ordered scalars. Reading from left to right, these form a word $w$ as explained in \fig{figmulti}. The $z_i$ is the internal scalar momenta to the right of each gluon $i$. \label{figmulti2}}
	\end{center}
\end{figure}

To write down the YM$\,+\,\phi^3$ amplitudes in an efficient manner, it is convenient to first construct a set of color orderings that will be repeatedly used in subsequent formulae. We will label the $k$ external gluons (or gravitons in YME) as $\{ 1,2,3, \ldots, k \}$ and the $m\ge2$ external scalars (or gluons in YME) as $\{k+1,k+2, \ldots, k+m \}$. From these sets, we construct the set of color orderings $\sigma_{123\cdots k}$, which will later be useful,
\bea
\alpha &=&  \{1,2,3, \ldots, k\}\,, ~~~~  \beta = \{k+2, \ldots, k+m-1\}\,,\no \\ 
\sigma_{123\cdots k} &=&  \Big\{\{k+1, \gamma, k+m\} \,\Big| \, \gamma \in \alpha \shuffle \beta \Big\}\, . \label{shuffle}
\eea
Note that $\sigma_{123\cdots k}$ is essentially the shuffle product between the gluon and scalar sets, except that we have separated out the first and last scalar, since they are always associated with a fixed position on the multiperipheral graph (this amounts to picking a subset of Kleiss-Kuijf-basis orderings). In other words, $\sigma_{123\cdots k}$ is the set of all permutations of all the particle labels such that gluons and scalars are strictly ordered among themselves, and $k+1$  $(k+m)$ is the first (last) element in each permutation. The size of these sets of color orderings is $|\sigma_{123\cdots k}| = (k+m-2)!/k!/(m-2)!$. The corresponding multiperipheral graphs are depicted in \fig{figmulti2}. 

In general, we can write the complete tree amplitude between $k$ gluons and $m\ge 2$ scalars as
\bea
&& {\cal A}^{\text{YM}+\phi^3}_{k,m}(1, \ldots, k \, |\, k+1,\ldots,k + m)  =-i g^{k+m-2} \lambda^{m-2}   \sum_{i \in \text{cubic}} {n_i  c_i \over D_i}  \hskip2cm   \label{scalaramp1} \\
&& \hskip4cm = -i g^{k+m-2} \lambda^{m-2}   \left[ \sum_{\, w \in \sigma_{12\ldots k}} \!\! \!\!
  n(w) A^{\phi^3}_{k+m}(w) ~ + ~ \text{Perm}(1, \ldots, k) \right] \no \\
 &&\hskip8 cm + ~\text{Perm}(k+2, \ldots, k+m-1)\,.  \no
\eea
$A^{\phi^3}_{k+m}(w)$ are amplitudes  in bi-adjoint $\phi^3$ theory that are color-ordered only with respect to one of the two colors, i.e. planar tree amplitudes built out of $\phi^3$ graphs that respect the ordering $w$ and have numerators $c_i$.\footnote{As an example, consider $A^{\phi^3}_4(1,2,3,4)= \frac{c_s}{s}+\frac{c_t}{t}$, with $c_s= f^{\hat a_1 \hat a_2 \hat x} f^{\hat x \hat a_3 \hat a_4}$, $c_t=f^{\hat a_1 \hat a_4 \hat x}f^{x a_3 a_2}$.} Although  the permutation sum is written differently, this formula is a DDM decomposition of ${\cal A}^{\text{YM}+\phi^3}_{k,m}$, except that the color factors and kinematic numerators have swapped roles (as allowed by color/kinematics duality). 

The numerators $n_i$ contain both kinematic and global flavor factors 
(the latter promoted to color factors in the YME theory),
\bea
n_i &=&  N_i\,  \widetilde C_i ~ + ~ \text{multi-trace terms}\,, \no \\  
n(w) &=& N(w) \, \widetilde C^{\rm DDM}(k+1,\ldots, k+m) ~ + ~ \text{multi-trace terms} \,,
\label{single_multi}
\eea
where $n(w)$ is used to denote the $n_i$ numerator that corresponds to a multiperipheral graph with ordering $w$. 
The factors $\widetilde C_i$ correspond to single-trace contributions of the global flavor group, of which $\widetilde C^{\rm DDM}(k+1,\ldots, k+m)$ is a string of structure constants obtained by removing the gluons in the word $w$ (since they are singlets of the global group) and dressing the rest of the interaction vertices with $F^{ABC}$s,
\be
\widetilde C^\text{DDM}(k+1, \ldots, k+m) = i^{m-2}\,   F^{ A_{k+1} A_{k+2}  X_1} F^{ X_1   A_{k+3}  X_2} \cdots F^{ X_{m-3}   A_{{k+m-1}} A_{k+m}}  \, .
\ee
The kinematic factors $N(w)$ in \eqn{scalaramp1} can thus be interpreted as the single-trace numerators of the multiperipheral graphs of the YM$\,+\,\phi^3$ theory after both the global and local group-theory factors have been stripped off.  Multi-trace terms that appear in \eqn{single_multi} are suppressed by powers of $1 / \lambda$, and we will leave them to future work.

Note that all the single-trace terms in the square bracket in \eqn{scalaramp1} are proportional to the $\widetilde C^\text{DDM}(k+1, \ldots, k+m)$ factor. The coefficient of this factor is the following flavor-ordered partial amplitude: 
\be
{\cal A}^{\text{YM}+\phi^3}_{k,m}(1, \ldots, k \, |\, k+1,\ldots,k + m)   =  - i  \sum_{\, w \in \sigma_{12\ldots k}} 
  N(w) A^{\phi^3}_{k+m}(w) ~ + ~ \text{Perm}(1, \ldots, k)  \, .
  \label{scalaramp2}
\ee
In the trace-basis decomposition, this partial amplitude is associated with the global-group trace factor ${\rm Tr}(T^{A_{k+1}} \cdots T^{A_{k+m}})$. Throughout the paper, calligraphic amplitudes like ${\cal A}^{\text{YM}+\phi^3}_{k,m}$ will indicate color-dressed amplitudes irrespective of the flavor-dressing.

From here on, let us denote the single-trace multiperipheral numerators with $k$ gluons by $N_k(w)$.  From studying the Feynman rules of the YM$\,+\,\phi^3$ theory, we can deduce that these can be written in the following local form
\begin{equation}
 N_k(w) =  \prod_{i=1}^k 2 (\varepsilon_{i} \cdot z_i(w)) + \text{contact terms} 
 \label{schematicnum}\,,
\end{equation}
where the contact terms are the contributions that are proportional to an inverse propagator of the multiperipheral diagram. The vector variable $z_i=z_i(w)$ is used to denote the momentum of the internal scalar line to which the gluon attaches. It can be defined through the momenta of the external particles as
\begin{equation}
z_i(w)= \mathop{\sum_{1 \le j \le l}}_{w_l=i} p_{w_j}\,,
\end{equation}
where the external momenta are defined to be incoming. It is the sum of the momenta of all the particles to the left of the $i$-th gluon on the multiperipheral graph, including the momentum $p_i$.\footnote{Note that the $z_i$ variable is isomorphic to the region momenta $x_i=\sum_{j=1}^i p_i$ that was used in ref. \cite{Stieberger:2016lng}, but here the subscript on $z_i$ refers to the ``name'' of the gluon rather than its position. This notation simplifies the presentation of the formulae in this paper.} The factor $2 (\varepsilon_{i} \cdot z_i(w))$ is precisely the Feynman vertex for a gluon-scalar-scalar interaction. Thus, to first approximation, $N_k(w)$ is the product of $k$ such independent factors, precisely as in the Feynman-diagram numerator. 
However, color/kinematics duality and gauge invariance demand that  we also add some terms of the form $(\varepsilon_{i} \cdot \varepsilon_{j}) \, p_l^2$ to this numerator. If we did not do this, such terms would never get generated by kinematic Jacobi relations, contrary to what is expected from the Feynman rules, and thus the amplitude would be incorrect.

In the following, we will construct several nontrivial examples of \textit{local} color/ kinematics-satisfying numerators $N_k(w)$. These numerators differ from the conventional Feynman-graph ones by correction terms that can be assigned to the contact terms in the expression above.  

\subsection{YME amplitudes}

In general, the complete YME tree amplitude with $k$ gravitons and $m\ge 2$ gluons can be written as the following double copy between YM$\,+\,\phi^3$ and YM,
\bea
&& {\cal M}^{\text{YME}}_{k,m}(1, \ldots, k \, |\, k+1,\ldots,k + m)  = - i\Big( \frac{\kappa}{4}\Big)^{k} g_s^{m-2}  \sum_{i \in \text{cubic}} {n_i  n_i^{\rm YM} \over D_i}  \hskip3cm  \label{YMEamp}  \\
&& \hskip4cm = -i  \Big( \frac{\kappa}{4}\Big)^{k} g_s^{m-2} \left[ \sum_{\, w \in \sigma_{12\ldots k}} \!\! \!\!
  n(w) A^{\rm YM}_{k+m}(w) ~ + ~ \text{Perm}(1, \ldots, k) \right] \no \\
 &&\hskip8 cm + ~\text{Perm}(k+2, \ldots, k+m-1)\,.  \no
\eea
This formula has the exact same structure as \eqn{scalaramp1}, except that we have performed the replacements $c_i \rightarrow n_i^{\rm YM}$, $A^{\phi^3}_{k+m}(w) \rightarrow A^{\rm YM}_{k+m}(w)$,  $g \rightarrow \kappa/4 $, and  $\lambda \rightarrow  4 g_s/\kappa$. Also, scalars have been promoted to gluons, gluons to gravitons, and the global flavor symmetry has been promoted to a local gauge symmetry.\footnote{The local gauge symmetry of each gauge theory plays no role in the YME amplitude; those color factors $c_i$ do not enter the double copy (\ref{YMEamp}).} These replacements give valid gravitational amplitudes by virtue of color/kinematics duality and as explained in sec. \ref{gaugeinv}.  As before, the $n_i$ and $n(w)$ are the numerators of the YM$\,+\,\phi^3$ theory. 

In complete analogy with \eqn{scalaramp1}, the color-ordered single-trace YME amplitudes are obtained from the expression inside the square bracket of \eqn{YMEamp}, after stripping off the color factor $\widetilde C^\text{DDM}(k+1,\ldots, k+m)$,
 \begin{equation}
 M^{\text{YME}}_{k,m}(1,\ldots,k \, | \, k+1, \ldots, k+m) \! = \!\! \!\! \sum_{w \in \sigma_{12\ldots k}} \!\! \!\!
  N_k(w) A^{\text{YM}}_{k+m}(w) +
\text{Perm}(1, \ldots,k)\,. \label{ST_YME}
 \end{equation}
Note that, in the trace-basis decomposition, this partial amplitude has the color coefficient  $-i (\kappa/4)^k \, g_s^{m-2}\, {\rm Tr}(T^{A_{k+1}} \cdots T^{A_{k+m}})$.

To complete the description of the single-trace YME tree amplitude, we need to compute the numerator functions $N_k(w)$  in the YM$\,+\,\phi^3$ theory. We adopt the following procedure for constructing $N_k(w)$ case-by-case for each multiplicity $k$: 
\begin{enumerate}
 \item We will assume that $N_k(w)$ is a homogeneous polynomial of degree $k$ in the following building blocks made out of Lorentz-invariant scalar products:\footnote{
 For readability and compactness of formulae, in the following we shall denote the scalar product of vectors $a$ and $b$ as $(ab)$ rather 
 than the usual $a\cdot b$.}
\begin{equation}
\big\{ \, (\varepsilon_i  z_i) \, , \ \
 (p_i  z_i) \, , \ \
 (\varepsilon_i \varepsilon_j) \, , \ \
  (\varepsilon_i  p_j) \, , \ \
   (p_i  p_j) \, \big\} \, , \qquad i,j= 1,\ldots, k \, .
 \end{equation}
Each polarization vector needs
to appear exactly once in every monomial (i.e. the numerator is multilinear in $\varepsilon_i$). Similarly, we assume that $N_k(w)$ is at most linear in each $z_i$ (e.g. this is true of the Feynman diagrams).
Note that we only consider the gluon momenta $p_i$, for  $i\le k$, as allowed external momenta in the Ansatz. The momenta of the external scalars only feature implicitly through the $z_i(w)$ variables. Similarly, we assume that the dependence on the ordering $w$ only appears in $z_i(w)$, and thus any rational-valued free coefficients that we use in the Ansatz can be taken to be independent of $w$. All together, this implies that the size of the Ansatz is fixed and finite even when the number of scalars $m$ approaches infinity. This is a crucial property that allows us to write  YME amplitudes for a fixed  number of gravitons $k$ and an arbitrary  number of gluons $m$. 
\item We take $N_k(w)$ in the form shown in \eqn{schematicnum}.
Numerators have a term coming from the cubic YM$\,+\,\phi^3$ Feynman graphs plus additional corrections.
Each additional term needs to be  a contact term.
We find that it is sufficient to
include contact terms that are proportional to inverse propagators of the form
\begin{equation}
2(p_i z_i) = z_i^2 - (z_i - p_i)^2  \, .
\end{equation}
\item The equations that we use to constrain the Ansatz are obtained from demanding that \eqn{ST_YME} is gauge/diffeomorphism invariant, i.e. the YME amplitude should vanish upon replacing one polarization vector with the corresponding momentum in the numerator $N_k(w)$,
\begin{equation}
\varepsilon_i \rightarrow p_i \, .
\end{equation}
This fixes the contact terms in $N_k(w)$, up to terms that cancel out in the permutation sum of \eqn{ST_YME}. Note that it is necessary to use the BCJ amplitude relations for $A^{\text{YM}}_{k+m}(w)$ when imposing gauge invariance (e.g. see appendix \ref{BCJappendix}). Otherwise, the $N_k(w)$ are over-constrained to the point that no solution exists, since a solution would demand that each $N_k(w)$ is separately gauge invariant. This constraint is usually  too severe  for a local function. While this procedure fixes the YME amplitudes completely, the expressions for individual numerators are not unique as, for example, one can add to the amplitude terms proportional
to the BCJ relations. This residual freedom can be used to find particularly simple expressions for $N_k(w)$.
\end{enumerate}
Explicit results are presented in the following subsections.

\subsection{Semi-recursive amplitudes with $k\leq 5$ gravitons}

To present compact expressions and to uncover additional structure, it is convenient to seek a recursive presentation for our numerators. We first introduce a short-hand notation for the Feynman vertex that corresponds to a gluon attaching to a scalar line
\begin{equation}
 u_i = 2 (\varepsilon_i z_i) \, ,
\end{equation}
and then write the numerators on a recursive form,
\begin{equation}
 N_k = N_{k-1} u_k + 2 (p_k z_k) {\cal Q}_k \, , \label{recursive1}
\end{equation}
with $N_0=1$. The first term $N_{k-1} u_k$ is by construction giving the right factorization limit $\sim  M^{\text{YME}}_{k-1,l} \frac{1}{P^2} M^{\text{YME}}_{1,m-l}$ for the amplitude, since $u_k$ is the numerator entering $M^{\text{YME}}_{1,m-l}$. The correction term is manifestly a contact term since $2(p_k z_k)$ can be expressed as a difference of inverse propagators. Still, ${\cal Q}_k$ is an unknown polynomial of degree-$(k-1)$, and thus the formula (\ref{recursive1}) is only partially recursive.

For example, with two and three external gravitons we employ the Ans\"{a}tze
\begin{eqnarray}
{\cal Q}_2 &=& a_0 (\varepsilon_1 \varepsilon_2)  \, ,\\
 {\cal Q}_3 &=& a_1 (\varepsilon_1 \varepsilon_2) (\varepsilon_3 p_1) +
 a_2 (\varepsilon_1 \varepsilon_2) (\varepsilon_3 p_2) + a_3 (\varepsilon_1 \varepsilon_3) (\varepsilon_2 p_1) + a_4 (\varepsilon_1 \varepsilon_3) (\varepsilon_2 p_3) + \no \\
 &&  a_5 (\varepsilon_2 \varepsilon_3) (\varepsilon_1 p_2)+ a_6 (\varepsilon_2 \varepsilon_3) (\varepsilon_1 p_3) + a_7 (\varepsilon_1 \varepsilon_2) u_3  + a_8 (\varepsilon_1 \varepsilon_3) u_2 + a_9 (\varepsilon_2 \varepsilon_3) u_1 \, , \no  
\end{eqnarray}
where $a_i$ are free parameters. Enforcing gauge invariance, and using the BCJ amplitude relations in appendix \ref{BCJappendix}, we find the compact solution: $a_0=a_8=a_9=1$, $a_2=2$ and for the remaining $a_i=0$.

Going to higher points, we introduce some additional notation defining the functions
\begin{align}
B^\mu_{1234} \equiv &\,   (\varepsilon_1 \varepsilon_2) (\varepsilon_3 \varepsilon_4) [p_3-  p_4]^\mu +[(\varepsilon_1 \varepsilon_3) (\varepsilon_2 \varepsilon_4)-(\varepsilon_1 \varepsilon_4) (\varepsilon_2 \varepsilon_3)]  p_2^\mu\,, \no \\ 
D_{1234} \equiv &\, -2 (p_{1}z_{1}) (\varepsilon_{1}\varepsilon_{2}) (\varepsilon_{3}\varepsilon_{4})+p_{1\mu}B^\mu_{1234}\,, \no \\
E_{12345}  \equiv &\,4(p_1 z_1)  (\varepsilon_1 p_2) (\varepsilon_2 \varepsilon_3) (\varepsilon_4 \varepsilon_5) +4 [(p_1 p_2) \varepsilon_{1}  -  (\varepsilon_1 p_2) p_{1}]_\mu  B^\mu_{2345} \, .
\end{align}
With this preparation, we are able to give very compact expressions for ${\cal Q}_k$ in case of $k \leq 5$ external gravitons (and any number of gluons),
\begin{align}
{\cal Q}_1= &\, 0\,, \no \\
{\cal Q}_2= & \, (\varepsilon_1 \varepsilon_2)\,, \no  \\
{\cal Q}_3= & \,2   (\varepsilon_3 p_2) {\cal Q}_2  + u_1 (\varepsilon_2 \varepsilon_3)  + u_2 (\varepsilon_1 \varepsilon_3)  \,, \no \\
{\cal Q}_4= & \,   2 (\varepsilon_{4}p_{3}) {\cal Q}_3 + 2  (\varepsilon_{4}p_{2}) u_3 {\cal Q}_2
+ u_1 u_2  (\varepsilon_{3}\varepsilon_{4})  
+  u_1 u_3 (\varepsilon_{2}\varepsilon_{4}) 
+  u_2 u_3 (\varepsilon_{1}\varepsilon_{4}) 
+D_{1234}\,, \no \\
{\cal Q}_5=& \,
2 (\varepsilon_5 p_4) {\cal Q}_4 + 2  (\varepsilon_5 p_3)  u_4 {\cal Q}_3 + 2  (\varepsilon_5 p_2) u_4 u_3 {\cal Q}_2  \no \\ &
+ u_1 u_2 u_3  (\varepsilon_4 \varepsilon_5)  + u_1 u_2 u_4   (\varepsilon_3 \varepsilon_5) +  u_1 u_3 u_4  (\varepsilon_2 \varepsilon_5) +   u_2 u_3 u_4  (\varepsilon_1 \varepsilon_5)  \no \\ &
+u_1 D_{2345} + u_2 D_{1345}+ u_3 D_{1245}+ u_4 D_{1235}  +E_{12345}\,.
\label{n4gravcomp}
\end{align}
Then, through three gravitons, we can write our numerators explicitly as
\begin{align}
N_1 =&2 (\varepsilon_1  z_1)\,, \no  \\
N_2 = &4 (\varepsilon_1  z_1) (\varepsilon_2  z_2)  + 2 (\varepsilon_1  \varepsilon_2) (p_{2}  z_{2})\,, \no \\
N_3 =  &\big[8 (\varepsilon_1  z_1) (\varepsilon_2  z_2)  + 4 (\varepsilon_1  \varepsilon_2) (p_{2}  z_{2})  \big] (\varepsilon_3  z_3)\,,  \no \\ 
&+ 4 (p_{3} z_{3})
\big[ (\varepsilon_3 p_2) (\varepsilon_1 \varepsilon_2)  + (\varepsilon_2 \varepsilon_3)   (\varepsilon_1  z_1)  + (\varepsilon_1 \varepsilon_3)  (\varepsilon_2  z_2)  \big]\,.
\end{align}
The numerator function entering the one-graviton amplitude does not have any additional contact terms. This makes it unique, and thus it is identical to the one obtained by Stieberger and Taylor using string-theory techniques \cite{Stieberger:2016lng}.

For the reader's convenience, we also spell out the expression for the four- and five-graviton numerators. The former is
\begin{align}
N_4& \!=\!
16 (\varepsilon_1 z_1) (\varepsilon_2 z_2) (\varepsilon_3 z_3) (\varepsilon_4 z_4)
\!+\!8(\varepsilon_{1}\varepsilon_{2})(\varepsilon_{3}z_{3})(\varepsilon_{4}z_{4})(p_{2}z_{2})
\!+\!8(\varepsilon_{2}\varepsilon_{3})(\varepsilon_{1}z_{1})(\varepsilon_{4}z_{4})(p_{3}z_{3}) \no \\ &
+8(\varepsilon_{1}\varepsilon_{3})(\varepsilon_{2}z_{2})(\varepsilon_{4}z_{4})(p_{3}z_{3})
+8(\varepsilon_{3}\varepsilon_{4})(\varepsilon_{1}z_{1})(\varepsilon_{2}z_{2})(p_{4}z_{4})
+8(\varepsilon_{2}\varepsilon_{4})(\varepsilon_{1}z_{1})(\varepsilon_{3}z_{3})(p_{4}z_{4}) \no \\ &
+8(\varepsilon_{1}\varepsilon_{4})(\varepsilon_{2}z_{2})(\varepsilon_{3}z_{3})(p_{4}z_{4})
+8(\varepsilon_{1}\varepsilon_{2})(\varepsilon_{3}p_{2})(\varepsilon_{4}z_{4})(p_{3}z_{3})
+8(\varepsilon_{1}\varepsilon_{2})(\varepsilon_{3}z_{3})(\varepsilon_{4}p_{2})(p_{4}z_{4}) \no \\ &
+8(\varepsilon_{2}\varepsilon_{3})(\varepsilon_{1}z_{1})(\varepsilon_{4}p_{3})(p_{4}z_{4})
+8(\varepsilon_{1}\varepsilon_{3})(\varepsilon_{2}z_{2})(\varepsilon_{4}p_{3})(p_{4}z_{4})
-4(\varepsilon_{1}\varepsilon_{2})(\varepsilon_{3}\varepsilon_{4})(p_{1}z_{1})(p_{4}z_{4}) \no \\ &
+8(\varepsilon_{1}\varepsilon_{2})(\varepsilon_{3}p_{2})(\varepsilon_{4}p_{3})(p_{4}z_{4})
-2(\varepsilon_{1}\varepsilon_{2})(\varepsilon_{3}\varepsilon_{4})(p_{1}p_{4})(p_{4}z_{4})
+2(\varepsilon_{1}\varepsilon_{2})(\varepsilon_{3}\varepsilon_{4})(p_{1}p_{3})(p_{4}z_{4}) \no \\ &
+2(\varepsilon_{1}\varepsilon_{3})(\varepsilon_{2}\varepsilon_{4})(p_{1}p_{2})(p_{4}z_{4})
-2(\varepsilon_{1}\varepsilon_{4})(\varepsilon_{2}\varepsilon_{3})(p_{1}p_{2})(p_{4}z_{4})\,.
\label{n4} \end{align}
The five-point numerator is given as
\begin{equation}
 N_5 = 2(\varepsilon_{5}z_{5})  N_{4} + 2 (p_5 z_5) {\cal Q}_5 \, ,
\end{equation}
with ${\cal Q}_5$ written out explicitly, as
\begin{align} 
{\cal Q}_5  & = 8 (\varepsilon_1 z_1) (\varepsilon_2 z_2) (\varepsilon_3 z_3) (\varepsilon_4 \varepsilon_5) 
 +   8 (\varepsilon_1 z_1) (\varepsilon_2 z_2) (\varepsilon_3 \varepsilon_5) (\varepsilon_4 z_4) 
 +   8 (\varepsilon_1 z_1) (\varepsilon_2 \varepsilon_5) (\varepsilon_3 z_3) (\varepsilon_4 z_4) \no \\ &
 +   8 (\varepsilon_1 \varepsilon_5) (\varepsilon_2 z_2) (\varepsilon_3 z_3) (\varepsilon_4 z_4) 
 +   8 (\varepsilon_1 \varepsilon_2) (\varepsilon_3 z_3) (\varepsilon_4 z_4) (\varepsilon_5 p_2) 
 +   8 (\varepsilon_1 z_1) (\varepsilon_2 \varepsilon_3) (\varepsilon_4 z_4) (\varepsilon_5 p_3) \no \\ &
 +   8 (\varepsilon_1 \varepsilon_3) (\varepsilon_2 z_2) (\varepsilon_4 z_4) (\varepsilon_5 p_3) 
 +   8 (\varepsilon_1 \varepsilon_2) (\varepsilon_3 p_2) (\varepsilon_4 z_4) (\varepsilon_5 p_3) 
 +   8 (\varepsilon_1 z_1) (\varepsilon_2 z_2) (\varepsilon_3 \varepsilon_4) (\varepsilon_5 p_4) \no \\ &
 +   8 (\varepsilon_1 z_1) (\varepsilon_2 \varepsilon_4) (\varepsilon_3 z_3) (\varepsilon_5 p_4) 
 +   8 (\varepsilon_1 \varepsilon_4) (\varepsilon_2 z_2) (\varepsilon_3 z_3) (\varepsilon_5 p_4) 
 +   8 (\varepsilon_1 \varepsilon_2) (\varepsilon_3 z_3) (\varepsilon_4 p_2) (\varepsilon_5 p_4) \no \\ &
 +   8 (\varepsilon_1 z_1) (\varepsilon_2 \varepsilon_3) (\varepsilon_4 p_3) (\varepsilon_5 p_4) 
 +   8 (\varepsilon_1 \varepsilon_3) (\varepsilon_2 z_2) (\varepsilon_4 p_3) (\varepsilon_5 p_4) 
 +   8 (\varepsilon_1 \varepsilon_2) (\varepsilon_3 p_2) (\varepsilon_4 p_3) (\varepsilon_5 p_4) \no \\ &
-   4 (\varepsilon_1 p_3) (\varepsilon_2 \varepsilon_5) (\varepsilon_3 \varepsilon_4) (p_1 p_2) 
 +   4 (\varepsilon_1 p_3) (\varepsilon_2 \varepsilon_4) (\varepsilon_3 \varepsilon_5) (p_1 p_2) 
 +   4 (\varepsilon_1 p_4) (\varepsilon_2 \varepsilon_3) (\varepsilon_4 \varepsilon_5) (p_1 p_2) \no \\ &
-   4 (\varepsilon_1 p_5) (\varepsilon_2 \varepsilon_3) (\varepsilon_4 \varepsilon_5) (p_1 p_2) 
 +   4 (\varepsilon_1 p_2) (\varepsilon_2 \varepsilon_5) (\varepsilon_3 \varepsilon_4) (p_1 p_3) 
-   4 (\varepsilon_1 p_2) (\varepsilon_2 \varepsilon_4) (\varepsilon_3 \varepsilon_5) (p_1 p_3) \no \\ &
-   4 (\varepsilon_1 p_2) (\varepsilon_2 \varepsilon_3) (\varepsilon_4 \varepsilon_5) (p_1 p_4) 
 +   4 (\varepsilon_1 p_2) (\varepsilon_2 \varepsilon_3) (\varepsilon_4 \varepsilon_5) (p_1 p_5) 
 +   4 (\varepsilon_1 p_2) (\varepsilon_2 \varepsilon_3) (\varepsilon_4 \varepsilon_5) (p_1 z_1) \no \\ &
-   4 (\varepsilon_1 \varepsilon_3) (\varepsilon_2 z_2) (\varepsilon_4 \varepsilon_5) (p_1 z_1) 
-   4 (\varepsilon_1 \varepsilon_2) (\varepsilon_3 z_3) (\varepsilon_4 \varepsilon_5) (p_1 z_1) 
-   4 (\varepsilon_1 \varepsilon_2) (\varepsilon_3 \varepsilon_5) (\varepsilon_4 z_4) (p_1 z_1) \no \\ &
-   4 (\varepsilon_1 \varepsilon_2) (\varepsilon_3 \varepsilon_4) (\varepsilon_5 p_4) (p_1 z_1) 
-   4 (\varepsilon_1 z_1) (\varepsilon_2 \varepsilon_3) (\varepsilon_4 \varepsilon_5) (p_2 z_2)
-   2 (\varepsilon_1 \varepsilon_5) (\varepsilon_2 \varepsilon_4) (\varepsilon_3 z_3) (p_1 p_2) \no \\ &
 +   2 (\varepsilon_1 \varepsilon_4) (\varepsilon_2 \varepsilon_5) (\varepsilon_3 z_3) (p_1 p_2)
 -   2 (\varepsilon_1 \varepsilon_5) (\varepsilon_2 \varepsilon_3) (\varepsilon_4 z_4) (p_1 p_2) 
 +   2 (\varepsilon_1 \varepsilon_3) (\varepsilon_2 \varepsilon_5) (\varepsilon_4 z_4) (p_1 p_2) \no 
 \\ 
& -   2 (\varepsilon_1 \varepsilon_4) (\varepsilon_2 \varepsilon_3) (\varepsilon_5 p_4) (p_1 p_2) 
 +   2 (\varepsilon_1 \varepsilon_3) (\varepsilon_2 \varepsilon_4) (\varepsilon_5 p_4) (p_1 p_2)
 -   2 (\varepsilon_1 \varepsilon_5) (\varepsilon_2 z_2) (\varepsilon_3 \varepsilon_4) (p_1 p_3)\no \\ &
  +   2 (\varepsilon_1 \varepsilon_4) (\varepsilon_2 z_2) (\varepsilon_3 \varepsilon_5) (p_1 p_3) 
 +   2 (\varepsilon_1 \varepsilon_2) (\varepsilon_3 \varepsilon_5) (\varepsilon_4 z_4) (p_1 p_3) 
 +   2 (\varepsilon_1 \varepsilon_2) (\varepsilon_3 \varepsilon_4) (\varepsilon_5 p_4) (p_1 p_3) \no 
 \\ &
  +   2 (\varepsilon_1 \varepsilon_3) (\varepsilon_2 z_2) (\varepsilon_4 \varepsilon_5) (p_1 p_4) 
 +   2 (\varepsilon_1 \varepsilon_2) (\varepsilon_3 z_3) (\varepsilon_4 \varepsilon_5) (p_1 p_4) 
-   2 (\varepsilon_1 \varepsilon_2) (\varepsilon_3 \varepsilon_4) (\varepsilon_5 p_4) (p_1 p_4) \no \\ &
-   2 (\varepsilon_1 \varepsilon_3) (\varepsilon_2 z_2) (\varepsilon_4 \varepsilon_5) (p_1 p_5) 
-   2 (\varepsilon_1 \varepsilon_2) (\varepsilon_3 z_3) (\varepsilon_4 \varepsilon_5) (p_1 p_5) 
-   2 (\varepsilon_1 \varepsilon_2) (\varepsilon_3 \varepsilon_5) (\varepsilon_4 z_4) (p_1 p_5) \no \\ &
-   2 (\varepsilon_1 z_1) (\varepsilon_2 \varepsilon_5) (\varepsilon_3 \varepsilon_4) (p_2 p_3) 
 +   2 (\varepsilon_1 z_1) (\varepsilon_2 \varepsilon_4) (\varepsilon_3 \varepsilon_5) (p_2 p_3) 
 +   2 (\varepsilon_1 z_1) (\varepsilon_2 \varepsilon_3) (\varepsilon_4 \varepsilon_5) (p_2 p_4) \no \\ &
-   2 (\varepsilon_1 z_1) (\varepsilon_2 \varepsilon_3) (\varepsilon_4 \varepsilon_5) (p_2 p_5) \,.
\label{Q5} 
\end{align}

\subsection{Towards six gravitons and beyond}

Extrapolating from the observed structure of ${\cal Q}_k$ through five gravitons, we can write  a refined Ansatz for ${\cal Q}_6$,
\begin{align}
{\cal Q}_6=& \,
2 (\varepsilon_6 p_5) {\cal Q}_5+2 (\varepsilon_6 p_4) u_5 {\cal Q}_4 + 2  (\varepsilon_6 p_3)   u_5 u_4 {\cal Q}_3 + 2  (\varepsilon_6 p_2) u_5 u_4 u_3 {\cal Q}_2   + u_1 u_2 u_3 u_4  (\varepsilon_5 \varepsilon_6) \no \\ &
 + u_1 u_2 u_3 u_5 (\varepsilon_4 \varepsilon_6) + u_1 u_2 u_4 u_5  (\varepsilon_3 \varepsilon_6)+  u_1 u_3 u_4 u_5 (\varepsilon_2 \varepsilon_6)+   u_2 u_3 u_4 u_5 (\varepsilon_1 \varepsilon_6) \no \\ &
+u_1 u_2 D_{3456} +u_1 u_3 D_{2456}+u_1 u_4 D_{2356} +u_1 u_5 D_{2346}+u_2 u_3 D_{1456} \no \\ &
+u_2 u_4 D_{1356} +u_2 u_5 D_{1346}+u_3 u_4 D_{1256} +u_3 u_5 D_{1246} +u_4 u_5 D_{1236}  \no \\ &
+u_1 E_{23456}+u_2 E_{13456}+u_3 E_{12456}+u_4 E_{12356}+u_5 E_{12346}  \no \\ &
+F_{123456}\,, 
\end{align}
where $F_{123456}$ is an unknown polynomial that does not include any terms proportional to $u_i=2 (\varepsilon_i z_i)$ for any $i$. That is, all the $u_i$ terms are accounted for in the explicitly-listed contributions (we have checked this using numerics). Furthermore, by considering the $z_i\rightarrow \infty$ limits, we have deduced that the leading terms in $z_i$ are as follows: 
\begin{align}
F_{123456}= &-8 (p_1 z_1)  (\varepsilon_1 p_2)  (\varepsilon_2 p_3)  (\varepsilon_3 \varepsilon_4) (\varepsilon_5 \varepsilon_6)+z_i\text{-independent}~(\varepsilon_i \varepsilon_j)^2~\text{terms} \no \\
&\null +8(p_1 z_1) (p_3 z_3) (\varepsilon_1 \varepsilon_2) (\varepsilon_3 \varepsilon_4) (\varepsilon_5 \varepsilon_6)+z_i\text{-linear}~(\varepsilon_i \varepsilon_j)^3~\text{terms}\,.
\end{align}
Note that the first term above is very similar to the first term in both $D_{1234}$ and $E_{12345}$, suggesting that these are also given by a recursive formula. Likewise, the first term on the second line of $F_{123456}$ has a similar structure.

In general, the structure of ${\cal Q}_k$ appears to be the following formula
\be
{\cal Q}_k= \, \sum_{i=1}^{k-1} 2 (\varepsilon_k p_{i}) {\cal Q}_{i} \! \! \prod_{j=i+1}^{k-1} \! u_j  ~ + \! \! \! \sum_{\rho \! \! \subseteq \{1,\ldots,k-1\} } \! \! \! \! \!  G_{\rho, k}  \prod_{j\in \rho^{\complement}  } u_j\,,
\ee
where $\rho$ runs over all non-empty (sorted) subsets of $\{1,\ldots,k-1\}$ and $\rho^{\complement}$ is the complement of that subset ($\rho^{\complement}$ can be empty). The functions $G_{\rho, k}$ for $\rho=(123\cdots)$ are defined up to six gravitons as
\bea
G_{1,k}&=&(\varepsilon_1 \varepsilon_k)\,, \no \\
G_{12,k}&=&0\,, \no \\
G_{123,k}&=&D_{123k}\,, \no \\
G_{1234,k}&=&E_{1234k}\,, \no \\
G_{12345,k}&=&F_{12345k}\,.
\eea
Presumably, for every $k\ge7$,  a new $u_i$-independent function $G_{12345\cdots,k}$ should appear. The Ansatz for this function should be simpler than the one of ${\cal Q}_k$, as it should be a degree-$(k-1)$ polynomial in the reduced set of building blocks  
\begin{equation}
\big\{ \, (p_i  z_i) \, , \ \
 (\varepsilon_i \varepsilon_j) \, , \ \
  (\varepsilon_i  p_j) \, , \ \
   (p_i  p_j) \, \big\} \, , \qquad i,j= 1,\ldots, k \, .
 \end{equation}
 As already indicated, the terms inside $D_{123k}$,  $E_{1234k}$ and $F_{12345k}$ appear to have a recursive structure themselves, suggesting that it may be possible to determine $G_{\rho, k}$ to any multiplicity $k$.

\section{Loop-level amplitudes \label{loopsec}}

The strategy of constructing color/kinematics-satisfying numerators 
for the YM$\,+\,\phi^3$ theory by using an Ansatz compatible with
factorization properties and requiring gauge invariance can in principle be extended to loop level. An important ingredient of the tree-level construction is to use the BCJ amplitude relations when checking gauge invariance, and thus at loop level we would expect that analogs of these relations will be useful. Indeed, a
gauge transformation of the YM$\,+\,\phi^3$ numerators (or, alternatively, a diffeomorphism transformation of the YME amplitude) has the effect of dressing  the graphs of the YM amplitudes with additional factors depending on loop 
and external momenta. Diffeomorphism invariance of the YME theory demands that these combinations cancel upon integration.

Amplitude relations of this type, with the additional factors being linear in momentum invariants, have been discussed recently at one 
loop  \cite{Boels:2011mn,Tourkine:2016bak,Hohenegger:2017kqy} and at two and higher loops \cite{Tourkine:2016bak}. While they are reminiscent  
of the fundamental BCJ relations from which tree-level color-ordered amplitudes relations can be derived, the integration appears to be essential at loop level since in general there is an ambiguity in defining a unique integrand for non-planar contributions. 

In this section, we will express the all-loop one-graviton $m$-gluon integrands for YME theories in the generic Jordan family in terms of the corresponding (s)YM integrands, and, in the process, argue for the existence of all-loop, all-multiplicity amplitudes relations. 
To this end, we need the all-loop color/kinematics-satisfying numerators of the $n$-scalar, one-gluon amplitude in a configuration which contains no internal gluons. This contribution is uniquely specified by collecting all terms proportional to the coupling factor $ g\, (g \lambda/2)^{m+2L-2}$, which are leading in $\lambda$ (and, correspondingly, leading in $g_s$ in the YME theories). 

Without the external gluon, numerators that are leading in $\lambda$ follow trivially from the $\phi^3$ part of the Lagrangian \eqref{YMscalar}. The addition of the external gluon is simply governed by its minimal coupling with the scalars, which is proportional to one power of momentum.  On dimensional grounds, no contact terms are possible since they would correspond to inverse propagators (two or more powers of momenta) dressing the cubic graphs, and thus the color/kinematics-satisfying numerators are simply given by the 
Feynman rules of \eqref{YMscalar}. The kinematic Jacobi relations are easily shown to hold off-shell for these numerators, as they translate into momentum conservation around a $\phi^3$ vertex labeled by the legs $(i,j,k)$,
\be
(\varepsilon \cdot p_i) V^{\phi^3}(i,j,k)+ (\varepsilon \cdot p_j) V^{\phi^3}(i,j,k)+ (\varepsilon \cdot p_k) V^{\phi^3}(i,j,k) =0\,,
\ee
since $p_i+p_j+p_k=0$ holds for the momenta entering a given vertex.

\begin{figure}[t]
      \centering
      \includegraphics[scale=0.60]{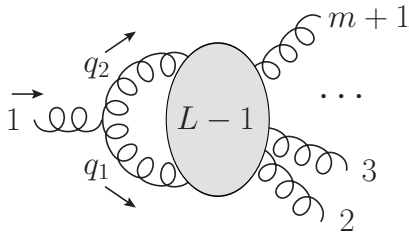}
      \caption[a]{\small A schematic representation of the $(m+1)$-point  $L$-loop YM amplitude where a cubic-graph vertex has been isolated. The shaded 
      blob stands for all the graphs contributing to this amplitude. In all such graphs, we uniformly name the internal momenta
      flowing in and out of the isolated vertex such that $p_1 = q_1+q_2$.
      \label{momentum_definition} }
\end{figure}

To express the result of the double copy in terms of YM amplitudes, we note that excising the graviton (i.e. removing the graviton and the vertex to which it is attached) yields in general $(m+1)$-gluon partial YM amplitudes of different trace structures (since the graviton is a singlet under the gauge group). Focusing on the contributions with the fewest number of traces, we note that we can restrict to the case of partial amplitudes that only involve single- or double-trace terms. 
We are therefore led to organize the single-graviton $L$-loop, $m$-gluon amplitudes in terms of the following YM $L$-loop $(m+1)$-gluon amplitude integrands of single- or double-trace type:
\begin{eqnarray}
&& 
A_{1\text{-trace}}^{{\rm YM},(L)}(2,\dots,i, 1, i+1,\dots m+1)
\quad \text{and}\quad
A_{2\text{-trace}}^{{\rm YM},(L)}(1 \,|\, 2,\dots, m+1)\,,
\label{12trace} 
\end{eqnarray}
where particle 1 is the gluon that is promoted to a graviton after the double copy is carried out.
Since we are talking about integrands, we must specify the loop momenta. We assume that, in a generic cubic-graph representation of the integrand, the internal loop momenta to the left and to the right of the first leg are $q_1$ and $q_2$ with $p_{1}=q_1+q_2$ (see \fig{momentum_definition}).
Note that the second amplitude in \eqn{12trace} does not contribute in SU($N_c$) YM because of the vanishing color factor ${\rm Tr}(T^{\hat a_1})$ and, for U($N_c$), the double trace will cancel against terms in the single trace because of $U(1)$ decoupling properties. However, this amplitude does a priori contribute to the YME amplitudes.\footnote{Similar cases of vanishing YM graphs which contribute nontrivially to the corresponding (super)gravity amplitudes occurred at four-loop 
four-point ${\cal N}=8$ supergravity amplitude \cite{Bern:2012uf} as well as at one-loop four-point amplitudes in ${\cal N}\le 4$ supergravities \cite{Carrasco:2012ca}. }

In terms of these partial amplitudes, the leading-$g_s$ single-trace contribution of the one-graviton $m$-gluon $L$-loop amplitude in a YME theory is given by 
\begin{eqnarray}
M_{1,m,L}^{\rm YME}(1\,|\, 2,3,\ldots,m+1) &=& A_{2\text{-trace}}^{{\rm YM},(L)}(1\, | \, 2,\dots , m+1)[2\, \varepsilon_{1}\cdot q_1]
\\
&&+ \hskip-0.3cm  \sum_{\text{cyclic}(2,\dots,m+1)}  \hskip-0.3cm  A_{1\text{-trace}}^{{\rm YM},(L)}(1,2,\dots, m+1) [2 \, \varepsilon_{1}\cdot q_1] \, ,
\nonumber
\end{eqnarray} 
where the notation $[\cdots]$ is used to emphasize that the factor in the bracket should multiply the integrand before performing the integration over $q_1$.

Following the general argument in sec.~\ref{gaugeinv}, this expression must be invariant under linearized diffeomorphisms, { i.e.} it should vanish
under the replacement $\varepsilon_{1}\rightarrow p_{1}$. It therefore follows that
\be
\label{eq_gi_Lloop}
0 = A_{2\text{-trace}}^{{\rm YM},(L)}(1 \, | \, 2,\dots, m+1)[p_{1}\cdot q_1]  
~~ 
- \hskip-0.3cm \sum_{\text{cyclic}(2,\dots,m+1)}  \hskip-0.3cm  A_{1\text{-trace}}^{{\rm YM},(L)}(1,2,\dots, m+1)  [ p_{1}\cdot q_1]     \, .
\ee
At one loop, this relation reproduces the field-theory limit of the result obtained 
by analyzing the  string-theory monodromy relations in ref. \cite{Hohenegger:2017kqy}.

A solution to this equation, which relies on a special parametrization of loop integrals, is that the two terms in \eqn{eq_gi_Lloop} 
vanish separately. It was shown in ref.~\cite{Bern:1990ux} that $U(1)$ decoupling implies
\be
0=A_{2\text{-trace}}^{{\rm YM},(L)}(1 \, | \, 2,\dots, m+1) ~~ + \hskip-0.3cm \sum_{\text{cyclic}(2,\dots,m+1)}  \hskip-0.3cm A_{1\text{-trace}}^{{\rm YM},(L)}(1,2,\dots, m+1) \, .
\ee 
While this equation holds for integrated partial amplitudes, one may consider {\it defining} the integrand of the double-trace amplitude to be the negative of the above sum of the integrands of the single-trace amplitudes. With such a choice, the two terms in \eqn{eq_gi_Lloop} become equal, and thus both vanish independently. Consequently, $L$-loop diffeomorphism invariance of the double-copy amplitude holds in this special 
parametrization if 
\be
 \sum_{\text{cyclic}(2,\dots,m+1)}  \hskip-0.3cm  A_{1\text{-trace}}^{{\rm YM},(L)}(1,2,\dots, m+1)  [ p_{1}\cdot q_1]   = 0 \, .
\ee
At one loop, this reproduces the form of the planar amplitudes relations argued for in refs.~\cite{Boels:2011mn} and \cite{Tourkine:2016bak}.

By considering multi-trace terms which are leading in $g_s$ in the YME theories, the internal loop gravitons are still suppressed by powers of $\kappa/g_s$, and similar relations can be derived as in \eqn{eq_gi_Lloop},
\begin{eqnarray}
0 &=& A_{(j+1)\text{-trace}}^{{\rm YM},(L)}(1 \, | \,  2,\dots \, | \,| \ldots  \, | \, \ldots , m+1)[p_{1}\cdot q_1]  
\\
&&- \hskip-0.3cm \mathop{\sum_\text{all distinct}}_{\text{insertions of}~1} A_{j\text{-trace}}^{{\rm YM},(L)}(2,\dots \, | \,| \ldots ,1, \ldots  \, | \, \ldots, m+1) [p_{1}\cdot q_1]     \,,
\nonumber
\end{eqnarray}
where $1 \le j \le {\rm min} (L,m)$. And similarly, a $U(1)$ decoupling identity can be used to define the $(j+1)$-trace integrand in terms of the $j$-trace integrands, such that the two above contributions vanish separately.

\section{Other theories \label{other}}

Although the main focus of this paper has been on YME amplitudes and numerators in the YM$\,+\,\phi^3$ theory, it is interesting to note that the results obtained for these cases can be recycled and used to construct tree amplitudes in a number of interesting theories. 
A particularly useful byproduct of the construction presented here is that
numerators of YM theory can be directly extrapolated from our results. 

\subsection{Pure YM numerators from YM$\,+\,\phi^3$ \label{otherYM} }
Starting from the $N_k(w)$ numerators for $(k+m)$-point amplitudes in YM$\,+\,\phi^3$ theory, we can consider the special case $m=2$, which corresponds to having a single scalar pair in the multiperipheral graph. 
The $\phi^3$ interaction does not contribute to these amplitudes, and hence the amplitude
\be
{\cal A}_{k+2}^{\rm sYM}(1,\ldots, k \,|\, k+1, k+2) ={\cal A}^{{\rm YM}+\phi^3}_{k,2}(1,\ldots, k \,|\, k+1, k+2)
\ee 
 is the same as the one in pure ${\cal N}=2$ or ${\cal N}=4$ sYM. This implies that supersymmetry Ward identities can be used to promote it to a pure-glue $(k+2)$-point amplitude (we will not spell out the Ward identities here). By linearity, the Ward identities can be applied directly on the $N_k(w)$ numerators to produce numerators in pure sYM. 

Alternatively, we can consider the factorization limit $P^2= (p_{k+1}+p_{k+2})^2\rightarrow 0$. The ${\cal A}^{{\rm YM}+\phi^3}_{k,2}$ amplitude should then factorize as
\be
\sum_{s\in \text{states}} {\cal A}^{\rm YM}_{k+1}(1,\ldots, k, P)^s \, \frac{1}{P^2} 2(\varepsilon_s(P) \cdot z_{k}) \equiv  -\frac{2}{P^2} (J_{\rm YM} \cdot z_{k}) \, ,
\label{factoriz}
\ee
where the second factor is the gluon-scalar-scalar vertex, $z_{k}=-p_{k+2}$ is the momentum of the last scalar, and the sum is over
all possible intermediate gluon states (labeled by their SO$(D-2)$ little group representations). We have written the factorization in terms of the amputated  on-shell  YM current $J_{\rm YM}$. This shows that, up to an overall factor, the $(k+1)$-point YM amplitude $(J_{\rm YM} \cdot \varepsilon(P) )$ is obtained by swapping $z_{k} \rightarrow \varepsilon(P)$ in \eqn{factoriz}. 

Using the kinematic Jacobi relations, we can construct the multiperipheral numerators of this $(k+1)$-point pure YM amplitude,
\be
n^{\rm YM}(w_1,w_2,\ldots, w_{k},P) \equiv \frac{1}{2}N_k(k+1,[[\cdots [[w_1,w_2],w_3],\ldots],w_k],k+2)\Big|_{z_k \rightarrow \varepsilon(P)}\,,
\label{factoriz2}
\ee
where the square brackets denote nested commutators, implying that the right-hand side is a sum over $2^{k-1}$ permuted numerators with associated signs weights coming from the commutators. Note that all the numerators on the right-hand side are exactly linear in $z_k$ because of \eqn{recursive1}, and hence the replacement $z_k \rightarrow \varepsilon(P)$ is consistent with linearity of the polarization vectors. Using the $N_k$ numerators constructed up to $k=5$, \eqn{factoriz2} gives explicit numerators in pure YM theory up to six points. 

Alternatively, we can directly obtain the $(k+2)$-point numerators in YM by dimensionally oxidizing the YM$\,+\,\phi^3$ theory so that the scalars become gluons. This is rather straightforward for a single scalar pair. The key insight is that, on dimensional grounds,   every term in the pure-glue YM numerators is proportional to at least one $(\varepsilon_{i} \cdot \varepsilon_{j})$ factor.  Picking out all terms proportional to a given $(\varepsilon_{i} \cdot \varepsilon_{j})$ factor  in the amplitude gives the amplitude in which legs $i$ and $j$ are scalars.  From this information we can reconstruct the pure-glue amplitude as a sum over all such terms, 
 \be
{\cal A}_{k+2}^{\rm YM}=\sum_{1\le i<j \le k+2}  (\varepsilon_{i} \cdot \varepsilon_{j}){\cal A}^{{\rm YM+\phi^3}}_{k,2}(1,\ldots \hat \imath, \ldots, \hat  \jmath, \ldots,k+2 \,|\, i, j)\Big|_{(\varepsilon_{a} \cdot \varepsilon_{b})^n \rightarrow \frac{1}{n}(\varepsilon_{a} \cdot \varepsilon_{b})^n}\,, 
\ee
where the hats on $\hat \imath$, $\hat \jmath$ mean that these legs are absent. Note that there would have been an overcounting of $(\varepsilon_{a} \cdot \varepsilon_{b})^n$ terms had we naively summed over all possible two-scalar amplitudes. However, this is easily fixed by introducing symmetry factors equal to $1/n$ for each such term. 

The same operation can be done at the level of numerators, giving
\be
n^{\rm YM}(w_1, \dots, w_{k+2})=\hskip-0.4cm  \sum_{1\le i<j \le k+2} \hskip-0.2cm (\varepsilon_{w_i} \cdot \varepsilon_{w_j}) \bar{N}_k (w_i,\alpha_i, w_{i+1}, \ldots, w_{j-1}, \beta_j, w_j)\Big|_{(\varepsilon_{a} \cdot \varepsilon_{b})^n \rightarrow \frac{1}{n}(\varepsilon_{a} \cdot \varepsilon_{b})^n }, 
\label{YMnum}
\ee
where $\alpha_i = [[\cdots[[w_1,w_2],w_3],\ldots],w_{i-1}]$, and $\beta_j = [w_{j+1},[\ldots, [w_{k},[w_{k+1},[w_{k+2}]]\cdots]]$ are nested commutators, which imply further summation with terms weighted by the sign of the commutator. Note that the summation range of $i,j$ implies that $\alpha_i$ and $\beta_j$ can contain one or zero elements (despite being commutators). In the first case we define $\alpha_2=w_1$,  $\beta_{k+1}=w_{k+2}$. For the second case, we have $\alpha_1=\beta_{k+2}=\emptyset$, with the additional caveat that each $\emptyset$ gives the numerator an overall $(-1)$ weight. Overall, taking into account the different terms in the commutators, the summation above has a total of $(k+2)2^{k-1}$ terms. The $\bar{N}_k$ are the left--right averaged multiperipheral YM$\,+\,\phi^3$ numerators, $\bar{N}_k(w)=(N_k(w_1,w_2,\dots, w_n) + (-1)^n N_k(w_n,w_{n-1},\dots, w_1))/2$. This reflection symmetry is not manifest for the scalar-gluon numerators, but for the pure-gluon numerators it is helpful to make it manifest.  

We now give some examples, starting with the three-gluon numerator
\be
n^{\rm YM}(2,1,3)=2(\varepsilon_{2} \varepsilon_{3}) (\varepsilon_{1} z_1) - (1 \leftrightarrow 2)- (1 \leftrightarrow 3) \, ,
\ee
where $z_1=p_1+p_2$ should be used when performing the permutation sum. For the four-point case, let us define
\bea
 \hat N_2(3,1, 2, 4)&=& (\varepsilon_{3} \varepsilon_{4})  \bar{N}_2(3,1, 2, 4)\Big|_{(\varepsilon_{i} \cdot \varepsilon_{j})^n \rightarrow \frac{1}{n}(\varepsilon_{i} \cdot \varepsilon_{j})^n} 
 \no \\ & =&(\varepsilon_{3} \varepsilon_{4}) \Big[4(\varepsilon_{1} z_1)(\varepsilon_{2} z_2)  +\frac{1}{2}(p_2 z_2)(\varepsilon_{1} \varepsilon_{2})-\frac{1}{2}(p_1 z_1)(\varepsilon_{1} \varepsilon_{2})\Big], 
\eea
where $z_1=p_1+p_3$, $z_2=p_1+p_2+p_3$. Note that the $(\varepsilon_{1} \varepsilon_{2}) (\varepsilon_{3} \varepsilon_{4})$ terms has been rescaled compared to the $N_2$ numerator, and the last term is there because of the left--right averaging.  Then the 
four-gluon numerator is
\be
n^{\rm YM}(3,1,2,4)= \hat N_2([3, 1], [2, 4])+ \hat N_2(2, [1, 3], 4)+\hat N_2(3, [4, 2], 1)\,,
\ee
where we for reasons of compactness have recombined several terms in the overall sum of (\ref{YMnum}) using the commutator notation.
Similarly, the five-gluon numerator is
\bea
n^{\rm YM}(4,1,2,3,5)&=& \hat N_3([[4,1],2],[3,5])-\hat N_3(3,[[4,1],2],5) \no \\ && -\hat N_3([4,1],[3,5],2)-\hat N_3(4,[2,[3,5]],1) \, , 
\eea
where
\bea
\hat N_3(4,1,2,3,5)&=&(\varepsilon_{4} \varepsilon_{5})  \bar{N}_3(4,1,2,3,5)\Big|_{(\varepsilon_{i} \cdot \varepsilon_{j})^n \rightarrow \frac{1}{n}(\varepsilon_{i} \cdot \varepsilon_{j})^n} \no \\
&=&\frac{1}{2}(\varepsilon_{4} \varepsilon_{5}) \Big\{\big[8 (\varepsilon_1  z_1) (\varepsilon_2  z_2)  + 2 (\varepsilon_1  \varepsilon_2) (p_{2}  z_{2})  \big] (\varepsilon_3  z_3)
\no \\ && \null \hskip 1.5cm + 2 (p_{3} z_{3})
\big[ (\varepsilon_3 p_2) (\varepsilon_1 \varepsilon_2)  + (\varepsilon_2 \varepsilon_3)   (\varepsilon_1  z_1)  + (\varepsilon_1 \varepsilon_3)  (\varepsilon_2  z_2)  \big]\Big\} \no \\
&& \null \hskip 3cm - \Big((4,1,2,3,5) \rightarrow (5,3,2,1,4)\Big) \,.
\eea
Here the last line comes from the left--right averaging, and it instructs the reader to subtract the same terms as above with permuted indices. Note that $z_i$ flips sign under this operation (e.g. $z_4 \rightarrow -z_5\,,\, z_1 \rightarrow -z_3$, etc.), since they are defined to point towards the right in the multiperipheral graph, see \fig{figmulti2}.

Using the numerators that we have constructed in this paper, and through use of \eqn{YMnum}, we thus obtain explicit numerators in pure YM up to seven points.

\subsection{Generalizations to DBI, NLSM and string theory}

With the explicit color/kinematics duality-satisfying numerators for the YM$\,+\,\phi^3$ theory derived in \sec{otherYM}, there are several other 
interesting double-copy amplitudes that can be constructed:

\begin{itemize}[leftmargin=15pt]
\item
The double copy $({\rm YM}+\phi^3)\otimes ({\rm YM}+\phi^3)$~\cite{Cachazo:2014nsa}, which gives non-supersymmetric Einstein gravity coupled to two gauge theories with different gauge groups, and scalars which transform in the bi-adjoint representation of these groups with  a $\phi^3$ self-coupling. The once-color-stripped amplitudes in the single-trace sector are given by
\begin{equation}
M^{{\rm GR}\,+\,{\rm YM}_1+\,{\rm YM}_2\,+\,\phi^3}_{k,m} = \sum_{w \in \sigma_{12\cdots k}} \!\! \!\!
  N_k(w) A^{\text{YM}+\phi^3}_{k+m}(w) +{\rm Perm}(1,\ldots,k) \, .
 \end{equation}

\item
Using the fact that the nonlinear sigma model (NLSM) obeys color/kinematics duality~\cite{Chen:2013fya}, the double copy $({\rm YM}+\phi^3) \otimes {\rm NLSM} $ gives amplitudes in the Dirac-Born-Infeld (DBI) theory\footnote{In this context we do not distinguish between DBI and BI theories.} coupled to the NLSM model~\cite{Cachazo:2014xea}.
The color-ordered single-trace DBI$\,+\,$NLSM amplitudes are obtained by replacing $A^{\rm YM}$ with $A^{\rm NLSM}$ in \eqn{ST_YME},
 \begin{equation}
 M^{{\rm DBI}\,+\,{\rm NLSM}}_{k,m}(1,\ldots,k \, | \, k+1, \ldots, k+m) \! = \!\! \!\! \sum_{w \in \sigma_{12\ldots k}} \!\! \!\!
  N_k(w) A^{\text{NLSM}}_{k+m}(w) +
\text{Perm}(1, \ldots,k)\,. \label{DBI_NLSM}
 \end{equation}
 However, since simple color/kinematics-satisfying tree-level numerators are known in NLSM  \cite{Du:2016tbc,Cheung:2016prv}, a 
 better way to write the complete DBI$\,+\,$NLSM amplitudes (including all color factors, and all multi-trace terms) is
 \begin{equation}
{\cal M}^{{\rm DBI}\,+\,{\rm NLSM}}_{n} = \sum_{w \in S_{n-2}} \!\! \!\!
  n^{\rm NLSM}(w) A^{\text{YM}+\phi^3}_{n}(w)\,,
 \end{equation}
 where $n^{\rm NLSM}(w)= (-1)^{n/2}\prod_{i=1}^n 2(p_i z_i)$ is obtained from ref.~\cite{Carrasco:2016ldy} or from the Feynman rules of ref.~\cite{Cheung:2016prv}, and we have for simplicity set the overall coupling constants to unity. 
 
 \item
 The construction can be further generalized by considering the theory NLSM$\,+\,\phi^3$ \cite{Cachazo:2016njl,Carrasco:2016ygv}, whose multiperipheral single-trace numerators are a direct generalization of the $n^{\rm NLSM}(w)$ numerators,
 \be
 N^{{\rm NLSM}\,+\,\phi^3}_k(w) =  (-1)^{k/2}\prod_{i=1}^k 2(p_i z_i) \, ,
 \ee
 where $k$ is the number of adjoint NLSM scalars, and the remaining $m$ scalars are bi-adjoint.
The color-ordered single-trace ${\rm DBI}\,+\,{\rm YM}\,+\,{\rm NLSM}\,+\,\phi^3$ amplitudes are obtained using the $({\rm YM}+\phi^3) \otimes ({\rm NLSM}+\phi^3)$ double copy,
\begin{equation}
M^{{\rm DBI}\,+\,{\rm YM}\,+\,{\rm NLSM}\,+\,\phi^3}_{k,m} = \sum_{w \in \sigma_{12\cdots k}} \!\! \!\!
  N^{{\rm NLSM}\,+\,\phi^3}_k(w) A^{\text{YM}+\phi^3}_{k+m}(w) +{\rm Perm}(1,\ldots,k) \, .
 \end{equation}

 \item The numerators for pure YM that we explicitly constructed up to seven points in \eqn{YMnum} can be used to rewrite string tree amplitudes in novel forms. For example, the single-trace open superstring amplitudes for massless external states can be expressed as
\be
A^\text{OSS}(\tilde w)= \sum_{w \in S_{n-2}}  Z(\tilde w | w) \, n^{\rm sYM}(w) \, .
\ee
$Z(\tilde w | w)$ are the disk integrals introduced by Br\"{o}del, Stieberger and Schlotterer~\cite{Broedel:2013tta},
\be
Z(\tilde w | w)= \frac{1}{{\cal V}_{ SL(2)}} \int_{z_{\tilde w_1} <  z_{\tilde w_2} < \ldots <  z_{\tilde w_n} } d^nz \, \frac{\prod_{i<j}^n |z_{ij}|^{\alpha' s_{ij}}}{z_{w_1w_2} z_{w_2w_3} \cdots z_{w_n w_1}} \,,
\ee
where $z_{ij}=z_i-z_j$ are coordinates on the disk boundary (not to be confused with the momenta $z_i$ introduced in previous sections). It follows from color/kinematics duality and the DDM decomposition that this formula is equivalent to the decomposition of open superstring amplitudes in terms of YM trees that was introduced in refs.~\cite{Mafra:2011nv, Mafra:2011nw}.

Similarly, from the same line of reasoning, one can obtain the closed IIA/B superstring amplitudes for massless external states as
\be
A^\text{CSS}= \sum_{\tilde w \in S_{n-2}}  \sum_{w \in S_{n-2}}  \tilde n^{\rm sYM}(\tilde w) \,  {\rm sv}\{ Z(\tilde w | w)\} \, n^{\rm sYM}(w) \, ,
\ee
where ${\rm sv}\{ Z(\tilde w | w)\}$ is the so-called single-value projection of the disk integrals introduced by Stieberger and Taylor~\cite{Stieberger:2014hba}. Note that this projection can be identified with sphere integrals as explained in ref.~\cite{Stieberger:2014hba}.  
The distinction between $ \tilde n^{\rm sYM}(\tilde w)$ and $n^{\rm sYM}(w)$ is that they could be numerators from either the ${\cal N}=(1,0)$ or ${\cal N}=(0,1)$ sYM theory in $D=10$, and in a non-supersymmetric treatment they could be numerators of different component amplitudes. In the current context, where we have access to pure-gluon YM numerators, there is no distinction between the two cases.  

\end{itemize}

\section{Discussion and outlook \label{discussion}}

Employing the double-copy construction for YME theories~\cite{Chiodaroli:2014xia}, the DDM decomposition~\cite{DelDuca:1999rs}, and gauge invariance, we derived simple explicit presentations of single-trace tree-level YME scattering amplitudes with any number of gluons and 
up to five external gravitons in $D$ dimensions.
Our expressions are  linear combinations of tree-level all-multiplicity single-trace amplitudes of YM theory in which the coefficients are YM$\,+\,\phi^3$ numerator factors. 
Replacing the YM amplitudes with superamplitudes with the desired amount of supersymmetry leads effortlessly to supergravity scattering amplitudes, such as ${\cal N}=4$ YME theories and ${\cal N}=2$ supergravities belonging to the generic Jordan family, or truncations thereof.

Our construction also yields the color/kinematics-satisfying numerator factors for the YM$\,+\,\phi^3$ theory with bi-adjoint scalars.
Each structure present in the numerator for $k$ external gluons  (or gravitons, from the perspective of the YME theory) recursively 
generates the corresponding structure for $k+1$ gluons. At the same time, additional  structures appear for each $k$. While their dependence on momenta and 
polarization vectors is relatively simple, their complete form is yet to be determined.  

Even at tree level, the construction of local numerator factors that manifestly obey color/kinematics duality for YM amplitudes is a difficult problem that requires a better understanding (see ref. \cite{Mafra:2011kj} for a pure-spinor approach).  A by-product of our construction is an algorithmic procedure for numerators in both YM$\,+\,\phi^3$ and pure YM theory. It goes as follows: (1) One constructs the amplitudes in a DDM decomposition for at least two external scalars and $k$ external gluons where the color factors have been replaced by unknown kinematic coefficients; (2) Using the BCJ amplitude relations and enforcing gauge invariance guarantee that the solution for the kinematic coefficients is a set of master numerators for the YM$\,+\,\phi^3$ theory, which in turn can be 
used to construct the remaining numerators through kinematic Jacobi relations; (3) pure YM numerators are obtained by transforming the last two scalars into gluons through identities which hold at tree level.

We have also given a simple expression for the $L$-loop $m$-gluon single-graviton amplitudes in the YME theories in terms of the corresponding YM amplitude integrands. Using diffeomorphism invariance of these amplitudes, we have derived an all-loop relation between single-trace (s)YM amplitudes which generalizes the tree-level fundamental BCJ relations and extends to all loop orders the one- and two-loop results of  \cite{Tourkine:2016bak} and \cite{Hohenegger:2017kqy}.

In this paper we have heavily relied on the invariance of amplitudes under gauge and diffeomorphism transformations. We have given a detailed argument that diffeomorphism invariance follows from gauge invariance of the two gauge theories that enter into the double copy construction. Thus, the double copy of two gauge theories obeying color/kinematics duality can always be made to gravitate.

While we do not treat the multi-graviton YME loop amplitudes in this paper, it is possible to use our tree expressions and the $Q$-cut representation of (one-)loop amplitudes \cite{Baadsgaard:2015twa} to construct relatively 
simple expressions, along the lines of ref. \cite{He:2016mzd}. Reorganizing the result in terms of standard Feynman integrals is, however, 
not completely straightforward. Carrying out our construction directly at one-loop  and comparing with the result of the $Q$-cut-generated expressions
suggests that, in the process of reorganizing the latter in terms of Feynman integrands, additional contact terms are generated whenever YM$\,+\,\phi^3$ gluons 
are adjacent. It would be desirable to derive general expressions for these contact terms.

We note that our YME formulae should, with only minor modifications,  extend to the Coulomb branch of the generic Jordan family of supergravities. A double-copy construction for these theories 
was identified in ref. \cite{Chiodaroli:2015rdg}. One gauge-theory factor entering this construction is the spontaneously-broken (s)YM theory which can also be realized as a dimensional reduction
of a higher-dimensional theory with the scalar VEV being related to non-zero momenta  in the compact dimension. The second gauge-theory factor is a YM$\,+\,\phi^3$ theory with its global symmetry explicitly broken to the desired supergravity gauge group. A key simplification is that the single-trace flavor sector of the explicitly broken YM$\,+\,\phi^3$ theory is also related to a higher-dimensional unbroken theory. For multi-trace contributions the identification with a higher-dimensional theory is somewhat more complicated as one needs to project out massive $W$ bosons that otherwise appear as intermediate states (in order to avoid massive gravitons in the multi-trace sector of YME). 
Thus, for the YME formulae in this paper, the single-trace tree-level graviton-gluon-$W$-boson scattering amplitudes on the supergravity Coulomb branch are obtained from those of the unbroken theory by dimensional reduction while identifying extra-dimensional momenta as follows: $p_i^M=(p_i^\mu, \pm m_i)$ for the external $W^{\pm}$ bosons and $p_i^M=(p_i^\mu, 0)$ for the external gravitons and massless gluons (internal particles automatically obtain correct masses through conservation of the extra-dimensional momenta). 

The generic Jordan family is one of  several families of ${\cal N}=2$ YME supergravities  which lift to five dimensions and thus are completely determined by their cubic couplings involving three vector fields \cite{Gunaydin:1984ak,
	Gunaydin:1985cu,Gunaydin:1986fg}. It would be particularly
interesting to explore the double-copy properties of some of the other families of theories and to provide a  construction for explicit tree- and loop-level scattering amplitudes. 
In particular, a large class of  Maxwell-Einstein theories whose scalar manifolds are homogeneous spaces has already been investigated from the double-copy perspective in ref.
\cite{Chiodaroli:2015wal}.

The construction of amplitudes in supergravities with gauged R-symmetry via the double-copy method remains an 
important open problem. The major obstacle is that, on the one hand, R-symmetry gauging introduces a potential whose 
critical points lead, in general, to ground states which are not Minkowski space; on the other hand, double-copy methods 
as developed so far require a translation-invariant ground state and a momentum-space S matrix.
Supersymmetric vacua of theories with gauged R-symmetry are in general anti-de Sitter. Hence, for these theories, 
one should study Witten diagrams, which are related to correlation functions of boundary operators, rather than scattering 
amplitudes. 
There exists, however, an infinite family of $U(1)_{\rm R}$-gauged ${\cal N} = 2$ Maxwell-Einstein and YME supergravities in 
five dimensions which exhibit Minkowski (non-supersymmetric) ground states  \cite{Gunaydin:1984ak,Gunaydin:2000xk}. 
These ground states survive dimensional reduction to four dimensions. Extending the double-copy construction to this family of theories 
would constitute an important first step towards uncovering its natural generalization to gauged supergravities  
in Anti-de Sitter spacetimes. We will address these issues in the near future.

{\bf Added note}: After the completion of this work, we became aware of ref. \cite{Fu:2017uzt}, which partly overlaps with our results.

\section*{Acknowledgements}

We thank Jan Plefka, Oliver Schlotterer, Stephan Stieberger, and Stefan Weinzierl for useful discussions.
M.C., H.J. and R.R. would like to thank the Mani L. Bhaumik Institute for Theoretical Physics at UCLA   
for hospitality during the workshop ``QCD Meets Gravity", where some of the results of this paper were presented.
H.J. thanks the
Mainz Institute for Theoretical Physics at Johannes Gutenberg University for hospitality during the workshop
``Amplitudes: practical and theoretical developments", where some of these results were presented.
The research of M.C. is supported in part by the Knut and Alice Wallenberg Foundation under grant KAW 2013.0235.
The research of H.J. is supported in part by the Swedish Research Council under grant 621-2014-5722, the Knut and 
Alice Wallenberg Foundation under grant KAW 2013.0235, and the Ragnar S\"{o}derberg Foundation under grant S1/16. 
The research of R.R. was also supported in part by the US Department of Energy under DOE Grant DE-SC0013699.

\appendix
\section{Novel presentation of the BCJ relations \label{BCJappendix}}

Here we give a novel presentation of the BCJ relations for tree level amplitudes, in a form that is well suited for the notation of building blocks used to construct numerators in this paper. 

Starting from the set of permutations $\sigma_{123\cdots n}$ introduced in \eqn{shuffle}, consider a summation of a function $X(w)$ over this set, and furthermore summing over the position of leg $1$ in $\sigma_{123\cdots n}$. We denote this set of summations by dropping the 1 in $\sigma_{123\cdots n}$,
\be
\sum_{w\in \sigma_{23\cdots n}} \!\! X(w) \equiv  \sum_{w\in\sigma_{123\cdots n}} \!\! X(w)  + \sum_{w\in\sigma_{213\cdots n}} \!\! X(w)  + \sum_{w\in\sigma_{231\cdots n}} \!\! X(w) +\ldots+ \sum_{w\in\sigma_{23\cdots n1}} \!\! X(w)\,,
\ee
where each term on the right-hand side is a permutation of the first term.
Repeating this summation for leg 2 we can define $\sigma_{3\cdots n}$, etc.

Using this notation we can write up BCJ relations on an unusual form. First note that the so-called fundamental BCJ relation takes the form
\be
0=\sum_{w \in \sigma_{1}} A^{\rm YM}(w) \,  (p_1 z_1)\,.
\ee
This is needed to show that the one-graviton amplitude is gauge invariant
\be
0=2 \sum_{w \in \sigma_{1}} A^{\rm YM}(w) \,  (\varepsilon_1 z_1)\Big|_{\varepsilon_1 \rightarrow p_1}\,.
\ee
For $k$ gravitons, we need more complicated generalizations of the BCJ relations to show gauge invariance. In the language of this paper, they can be conveniently written as
\bea
&&0=\sum_{w \in \sigma_{1234 \cdots k}} A^{\rm YM}(w) \,  \Big[(p_1  z_1)+(p_2  z_2)+ (p_3  z_3)+ (p_4  z_4)+ \ldots +(p_k  z_k)\Big]\,,  \no \\ &&
0=\sum_{w \in \sigma_{234\cdots  k}} A^{\rm YM}(w) \, f(y_1) \,  \Big[(p_2  z_2)+ (p_3  z_3)+ (p_4  z_4)+ \ldots +(p_k  z_k)\Big]\,,  \no \\ &&
0=\sum_{w \in \sigma_{34\cdots  k}} A^{\rm YM}(w) \, f(y_1, y_2) \,  \Big[(p_3  z_3)+ (p_4  z_4)+ \ldots + (p_k  z_k)\Big]\,,   \no \\ &&
0=\sum_{w \in \sigma_{4 \cdots k}} A^{\rm YM}(w) \, f(y_1, y_2, y_3) \,  \Big[(p_4  z_4)+ \ldots + (p_k  z_k)\Big]\,,   \no \\ && 
~~~\vdots  \no \\ && 
0=\sum_{w \in \sigma_{k}} A^{\rm YM}(w) \, f(y_1, \ldots, y_{k-1}) \,  (p_k  z_k)\,. 
\eea
Here $y_i=y_i(w)$ is the region momenta given by adding the momenta of all the gluons coming before graviton $i$ in the word $w$
\be  
y_i(w)= \mathop{\mathop{\sum_{1 \le j < l}}_{w_l=i }}_{w_j \not \in  \{1,\ldots, k\} } p_{w_j}\,.
\ee
This excludes the momenta of any gravitons appearing before $i$ (note: ``gravitons'' refers to the YME amplitude).
Finally, the functions $ f(y_1, y_2, \ldots, y_r)$ can depend on the graviton momenta and polarizations in any way, but the gluon momenta can only 
enter through the $y_i$ as indicated by the arguments. They allow us to compactly summarize all the YM amplitude relations that are relevant for 
verifying the gauge invariance of YME amplitudes. 
Note that these relations can in principle be obtained from repeated use of the fundamental BCJ relation.

\bibliographystyle{JHEP}
\bibliography{litCGJR}

\end{document}